\documentclass[lettersize,journal]{IEEEtran}
\usepackage{cite}
\usepackage{amsmath,amssymb,amsfonts}
\usepackage{amsthm}
\usepackage{algorithm}
\usepackage{algpseudocode}
\usepackage{graphicx}
\usepackage{textcomp}
\usepackage{xcolor}
\usepackage{hyperref}
\usepackage{booktabs}
\usepackage{multirow}
\usepackage[table]{xcolor} 
\usepackage{colortbl}
\usepackage{subfig, siunitx, adjustbox}
\usepackage{array}
\usepackage{url}
\usepackage{adjustbox}
\usepackage{verbatim}
\usepackage{graphicx}
\usepackage{cite}
\newtheorem{theorem}{Theorem}

\begin{document}

\title{End-to-End Mobility-Aware Multi-RIS Optimization via Blockage Detection and Closed-Form Riemannian Updates}

\author{Sehyun Ryu, Seungmin Choi, Hyun Jong Yang,~\IEEEmembership{Senior Member,~IEEE}, \\and John M. Cioffi,~\IEEEmembership{Life Fellow,~IEEE}

\thanks{Corresponding author: Hyun Jong Yang and John M. Cioffi }
\thanks{S. Ryu is with the Department of Electrical Engineering, Pohang University of Science and Technology, Pohang, Republic of Korea, and also with the Institute of New Media and Communications, Seoul National University, Seoul, Republic of Korea (e-mail: sh.ryu@postech.ac.kr).}
\thanks{S. Choi and H. J. Yang are with the Department of Electrical and Computer Engineering, Seoul National University, Seoul, Republic of Korea. H. J. Yang is also with the Institute of New Media and Communications, Seoul National University (e-mail: 
seungminchoi22@snu.ac.kr; hjyang@snu.ac.kr).}
\thanks{J. M. Cioffi is with the Department of Electrical Engineering, Stanford University, Stanford, CA 94305 USA (e-mail: cioffi@stanford.edu).}
}

\markboth{Submitted to IEEE Transactions on Wireless Communications}{Ryu \MakeLowercase{\textit{et al.}}: Blcokage...}


\maketitle

\begin{abstract}
Millimeter-wave (mmWave) multi-user MIMO systems are highly susceptible to dynamic blockages, and reconfigurable intelligent surfaces (RIS) have been introduced as a remedy.
However, RIS links can themselves be blocked, while existing studies often assume ideal availability.
This paper proposes an end-to-end mobility-aware multi-RIS optimization framework that integrates per-RIS blockage detection with closed-form Riemannian updates.
The base station transmits short indexed synchronization signals, enabling each user to identify blocked panels via a simple energy test.
Based on the detected feasible sets, we jointly optimize the BS precoder and RIS phases using a Stochastic Closed-form Riemannian Phase Alignment (SCRPA) algorithm, which ensures unit-modulus feasibility, monotone convergence, and low complexity.
Extensive simulations validate reliable blockage detection and demonstrate significant weighted sum-rate and scalability gains compared to existing baselines.
\end{abstract}



\section{Introduction}
Millimeter-wave (mmWave) communications are a cornerstone of future wireless systems due to the abundant spectrum above 24~GHz \cite{RAPPAPORT2019, SHAFI2017}, and 3GPP standards continue to migrate toward higher carrier frequencies \cite{TS138104}.
At these bands, diffraction is weak and propagation is largely line-of-sight (LoS) dominated, making links highly vulnerable to human and object-induced blockage, which can cause abrupt SNR degradation and service
interruptions \cite{JAIN2019BLCK, VIRK2020}.
Reconfigurable intelligent surfaces (RIS) have therefore attracted considerable attention as a means to reshape wireless propagation by controllably reflecting incident signals toward intended receivers \cite{RIS_WU2019, RIS_WU2020, RIS_RENZO2020}.
Metasurface-based RIS panels offer programmable per-element phase control, enabling coverage extension, interference management, and virtual LoS creation without active RF chains \cite{PROG_CHEN2020, PROG_WANG2022, LIANG2024FILTR}.

\textbf{Prior work on RIS optimization.}
Early RIS studies focused on joint optimization of the BS precoder and RIS phase matrix to minimize transmit power under quality-of-service (QoS) constraints \cite{RIS_WU2019, TX_SUN2020, TX_YANG2021}, followed by extensions incorporating explicit user-level QoS balancing \cite{TXQ_LIU2021, TXQ_OBEED2022, TXQ_HE2023}.
Parallel efforts examined RIS placement, orientation, and deployment, showing that proper geometry can significantly enhance effective channel strength \cite{PLACE_BIE2022, PLACE_CHENG2022}.
As the field matured, \emph{weighted sum-rate} (WSR) maximization emerged as the dominant objective in multi-user (MU) systems.
For MU-MISO downlink, fractional programming, alternating optimization (AO), and manifold-based methods have been widely used to address the nonconvex coupling between active precoding and passive phase control
\cite{WSR_MUMISO_GUO2020, WSR_MUMISO_BAHINGAYI2025, WSR_mmW_FPAO_DAMPAHALAGE2021, WSR_mmW_FPAO_MA2023}.
The work of \cite{SADIA2025WSR} presents a phase-cooperative WSR design based on AO, where a branch-reduce-and-bound (BRB) global search computes the active beamformers and semidefinite relaxation (SDR) optimizes the RIS phases under perfect CSI and quasi-static channels.
Practical hardware constraints, notably discrete or quantized RIS phase shifts, have also been incorporated within similar AO frameworks \cite{WSR_Q_DI2020, WSR_Q_ABEYWICKRAMA2020, WSR_Q_GUO2024}.
Nonlinear triangular multiuser transceivers, such as dirty-paper coding (DPC)~\cite{COSTA1983DPC, YUCIOFFI2004DPC} and generalized decision-feedback equalization (GDFE)~\cite{CIOFFI1997GDFE}, fall within the canonical multiuser transmission framework and are capacity-achieving under perfect instantaneous CSIT.
Nevertheless, the RIS-aided WSR literature has primarily developed within the linear precoding framework, which readily integrates with practical considerations such as quantized phase control, imperfect CSIT, and real-time optimization.

\textbf{Multi-RIS systems and prevailing assumptions.}
To further extend coverage and bypass local obstructions, single-RIS designs have been generalized to \emph{multi}-RIS deployments with multiple cooperative panels.
A large portion of this literature still converges on WSR maximization under coupled active–passive constraints, spanning MU-MIMO/MU-MISO beamforming, synchronization, and channel estimation \cite{DRIS_HAN2020, DRIS_YOU2021, DRIS_DING2024, DRIS_ZHENG2021, NIU2022}.
Related RIS designs—such as filtering~\cite{LIANG2024FILTR, ALDABABSA2023MRIS}, hybrid precoding~\cite{LAN2024PHASE}, ergodic-rate analysis~\cite{WANG2024ERGO}, non-orthogonal multiple access (NOMA)~\cite{WANG2023MRIS}, and learning-based control \cite{SHARMA2024OL}—adopt AO-type frameworks that target WSR or its tight surrogates.
Additional directions, such as cooperative wideband OFDM \cite{HE2021CMRIS}, federated-learning–aided joint design \cite{NI2022IOTJ}, age-of-information–centric sensing \cite{QU2024OIF}, low-complexity beam training \cite{XU2024LCBT}, information-theoretic analyses \cite{ZHENG2023MI, LI2024MRISMU}, and coverage/deployment optimization \cite{MA2024MRIS}, complement this rate-centric view.
Nevertheless, foundational multi-RIS formulations typically assume perfect or quasi-static channel state information (CSI) during optimization \cite{LI2020, LI2024WSRCRA, SAEIDI2021}, an assumption that becomes increasingly fragile at mmWave frequencies under mobility.

\textbf{Blockage-aware designs and open challenges.}
Since RIS is fundamentally motivated by blockage mitigation, several studies have examined blocked RIS links.
Early work optimized beamformers under randomly blocked channels \cite{KUMAR2021}, while later efforts employed statistical or learning-based blockage models to design robust precoders \cite{JIAO2022, ZHOU2021}.
More recently, \cite{SARKER2024} optimized WSR for mmWave massive MIMO with mobile RIS under \emph{known} blockage states.
Despite this progress, two critical gaps remain.
First, per-RIS blockage must be detected \emph{online} with negligible overhead in multi-RIS systems, where superposed multipath components from different panels make RIS-wise isolation nontrivial.
Second, the subsequent joint optimization must complete within each \emph{CSI update interval}, which is configured according to the CSI reporting and reference signal periodicity specified in the NR standard \cite{TS38331, TS38214}.
Although \cite{YANG2024} proposed a Neyman–Pearson (NP)-based LoS blockage detector for a single-RIS system, extending this framework to multi-RIS scenarios becomes challenging due to inter-panel interference and additional synchronization overhead.
To the best of our knowledge, real-time multi-RIS optimization under time-varying channels has not been systematically addressed.

\textbf{Our approach and contributions.}
This work proposes an end-to-end framework for \emph{blockage-aware WSR maximization} in multi-RIS downlink systems under user mobility.
This work is an extended version of our GLOBECOM 2026 paper and is available on arXiv~\cite{RYU2026RIS}.
Compared with that version, this paper incorporates a standard $m$-sequence-based blockage detection scheme and develops a mobility-aware end-to-end multi-RIS optimization method under 3GPP-compliant dynamic scenarios.
The main contributions are summarized as follows:

\begin{itemize}
\item[\textbf{1)}] 
\textbf{Standard-compliant per-RIS blockage detection with negligible overhead.}
A lightweight blockage detection mechanism in which the BS periodically transmits short, \emph{standard-compliant} per-RIS indexed synchronization sequences, implements cyclic shifts of $m$-sequences aligned with 3GPP NR \cite{TS138211}.
Each UE performs a simple per-index correlation/energy test to classify RIS panels as blocked or unblocked, yielding an \emph{active RIS set} without learning, environment replay, or signaling overhead.

\item[\textbf{2)}] 
\textbf{Joint BS--RIS optimization under mobility-induced CSIT imperfection.}
Given the detected active RIS set, the joint optimization of the BS precoder and the quantized RIS phases under \emph{time-varying and imperfect CSIT} explicitly accounts for feedback delay and channel aging. 
This formulation captures practical mobility effects that existing RIS-aided WSR designs often ignore. 
In line with prior RIS-aided WSR studies, this work adopts linear precoding for real-time RIS control under imperfect CSIT, while multiple RISs can increase the angular and spatial diversity of the effective channel, improving user separability for multiuser transmission.

\item[\textbf{3)}] 
\textbf{Fast stochastic MM-based optimization with quantized RIS updates.}
A scalable real-time operation introduces \emph{Stochastic Closed-form Riemannian Phase Alignment (SCRPA)}, a block majorization--minimization (MM) algorithm \cite{RAZAVIYAYN2013BMM}. 
SCRPA admits closed-form BS updates via WMMSE/Lagrangian optimization and elementwise unit-modulus-preserving RIS updates with in-place $B$-bit quantization. 
Through timing and complexity analysis that the resulting detection--optimization loop fits within the NR CSI update interval for typical mmWave carrier frequencies and mobility profiles.
\end{itemize}

Ray-tracing-based simulations using NVIDIA Sionna \cite{NVIDIA} demonstrate reliable per-RIS blockage detection, robustness to CSIT aging and feedback delay, and improved WSR and convergence efficiency compared to representative AO and manifold-based baselines \cite{LI2020, WSR_MUMISO_GUO2020, NIU2022, LI2024WSRCRA}.

\textbf{Organization.}
Section~\ref{sec:system_model} presents the system model and problem formulation.
Sections~\ref{sec:blockage_detection} and \ref{sec:scrpa} describe the blockage detection, robust WSR optimization, and SCRPA algorithm.
Sections~\ref{sec:experiments} evaluate complexity, robustness, and performance, followed by conclusions in Section~\ref{sec:conclusion}.

\begin{figure*}
\centering
    \includegraphics[width=0.8\linewidth]{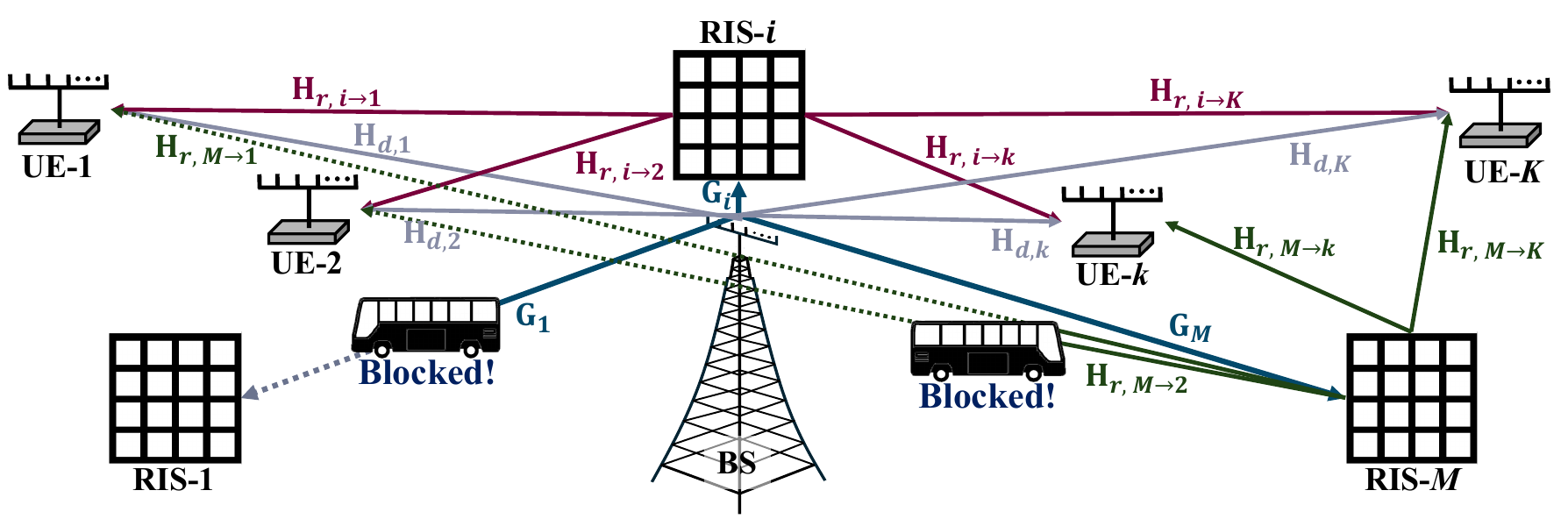}
    \caption{System model of the multi-RIS assisted mmWave MU-MIMO downlink system. The BS serves $K$ UEs with the aid of $M$ distributed RIS panels, where blockage effects limit the subset of RISs visible to each UE.}
    \label{fig1}
\end{figure*}

\section{System Model and Problem Formulation}
\label{sec:system_model}

\subsection{CP-OFDM Downlink and Multi-RIS MU--MIMO Model}
The model considers a downlink mmWave MU--MIMO system aided by $M$ distributed passive RIS panels and serving $K$ UEs.
The BS is equipped with $N_t$ transmit antennas and UE-$k$ has $N_{r,k}$ receive antennas.
RIS-$i$ consists of $N_i$ passive reflecting elements whose responses are controlled \emph{solely through phase shifts}.
Accordingly, the RIS control vector is $u_i\in\mathbb{C}^{N_i}$ with unit-modulus entries,
$|[u_i]_m|=1$, reflecting the absence of active amplification or baseband processing at the RIS.
Let $\mathcal{M}_{k}\subseteq\{1,\ldots,M\}$ denote the (unknown) set of RIS panels whose cascaded BS$\!\to\!$RIS$\!\to\!$UE-$k$ paths remain unblocked during the current beam-coherence interval.

\textbf{CP-OFDM and per-subcarrier notation.}
This model adopts a CP-OFDM downlink and presents all signals and channels \emph{per subcarrier}.
Each OFDM symbol prepends with a cyclic prefix (CP) of length $L_{\mathrm{cp}}$ samples.
$L_{\mathrm{cp}}$ exceeds the maximum delay spread among all direct BS$\!\to$UE and cascaded BS$\!\to$RIS$\!\to$UE links, such that after CP removal the linear convolution reduces to a circular one over the useful symbol duration.
This consequently estimates inter-symbol interference (ISI) and decouples the frequency-domain subcarriers, with each subcarrier experiencing an equivalent flat-fading MU--MIMO channel.
All subsequent expressions use this standard per-subcarrier equivalent model.
If residual ISI or inter-carrier interference (ICI) remains due to short CP length or phase noise, its effect absorbs the receive covariance matrix $\mathbf{C}_k$ as additional colored disturbance; this work focuses on the CP-sufficient regime.

\subsection{Synchronization and Effective Channel}
\label{subsec:synch_heff}

\subsubsection{Per-RIS Indexed Synchronization with CP}
\label{subsubsec:indexed_sync}
To tag RIS panels with negligible overhead, the BS assigns RIS-$i$ a cyclic shift $\nu_i=i-1$ of a length-$\ell$ $m$-sequence $s[n]$ generated by a primitive degree-$p$ linear feedback shift register (LFSR) ($\ell=2^p-1$) \cite{GOLOMB1967LFSR}:
\begin{equation}
s_i[n]\triangleq s[(n+\nu_i)\bmod \ell],\qquad M\le \ell.
\end{equation}
Such $m$-sequences have constant modulus and periodic correlations
$\sum_{n=0}^{\ell-1}s_i^{\!*}[n]s_j[n]=\ell$ if $i=j$ and $-1$ otherwise (cross-bias $-1/\ell$).
During sensing, the CP ensures that multipath convolution over the $\ell$ useful samples is circular, thereby preserving the above periodic correlation structure up to small residual leakage.

Unless stated otherwise, these $m$-sequences use $\ell=127$ (NR PSS-aligned), while longer LFSR lengths (e.g., $255,511$) are feasible when $M$ is large.
During a short sensing interval, the BS transmits the superposition
\begin{equation}
\mathbf{x}[n]=\sum_{i=1}^{M}\mathbf{v}_i s_i[n], 
\qquad 
\sum_{i=1}^{M}\|\mathbf{v}_i\|_2^2\le P_{\mathrm{pilot}},
\end{equation}
where $\mathbf{v}_i$ is a probing/steering vector toward RIS-$i$.
After CP removal and FFT (per subcarrier), UE-$k$ applies a fixed combiner; the model uses equal-gain combining
$\tilde{y}_k[n]=\tfrac{1}{\sqrt{N_{r,k}}}\mathbf{1}^H\mathbf{y}_k[n]$,
and forms the correlation output for RIS-$i$:
\begin{equation}
Z_{k,i}=\sum_{n=0}^{\ell-1}s_i^{\!*}[n]\tilde{y}_k[n].
\end{equation}

\textbf{Pilot-domain inter-index interference vs.\ data-domain MU interference.}
Two distinct types of interference arise and are treated separately:
(i) \emph{pilot-domain inter-index interference}, i.e., deterministic coupling among $\{Z_{k,i}\}$ caused by imperfect cyclic orthogonality under multipath and by superposition across RIS indices; this effect is explicitly modeled and mitigated (via whitening) in Section~\ref{sec:blockage_detection};
(ii) \emph{data-domain multi-user (MU) interference} during payload transmission, captured by the standard interference covariance in \eqref{eq:Rk_def}.
Pilot-domain coupling is a sensing-stage phenomenon and should not be confused with MU interference in the data phase.

\subsubsection{Effective Per-Subcarrier Channel}
\label{subsubsec:heff}
Let $\mathbf{H}_{d,k}\in\mathbb{C}^{N_{r,k}\times N_t}$ be the per-subcarrier BS$\to$UE-$k$ channel, 
$\mathbf{G}_i\in\mathbb{C}^{N_i\times N_t}$ the per-subcarrier BS$\to$RIS-$i$ channel, and 
$\mathbf{H}_{r,i\to k}\in\mathbb{C}^{N_{r,k}\times N_i}$ the per-subcarrier RIS-$i\to$UE-$k$ channel.
For an active-RIS set $\mathcal{S}_k\subseteq\{1,\dots,M\}$ and RIS phase vectors $\{u_i\}$ with $|[u_i]_m|=1$, the equivalent per-subcarrier downlink channel toward UE-$k$ after CP removal and FFT is
\begin{equation}
\label{eq:Heff_compact}
\mathbf{H}_{k,\mathrm{eff}}(\{u_i\};\mathcal{S}_k)
=
\mathbf{H}_{d,k}
+
\sum_{i\in\mathcal{S}_k}
\mathbf{H}_{r,i\to k}\,\mathrm{diag}(u_i)\,\mathbf{G}_i.
\end{equation}

\subsubsection{Imperfect CSIT with Per-Link Aging and Residual Uncertainty}
\label{subsubsec:imperfect_csit}
At sensing time $t_0$, the BS acquires noisy estimates 
$\widehat{\mathbf{H}}_{d,k}$, 
$\widehat{\mathbf{H}}_{r,i\to k}$, 
and $\widehat{\mathbf{G}}_i$.
Due to the control/feedback delay $\Delta$, the true channels at $t_0{+}\Delta$ deviate from their estimates.
Rather than aging the entire cascaded channel using a single scalar, the model uses first-order temporal correlation on a per-link basis:
\begin{align}
& \mathbf{H}_{d,k}^{\text{pred}}
= \rho_{d,k}(\Delta)\,\widehat{\mathbf{H}}_{d,k}, \\
& \mathbf{H}_{r,i\to k}^{\text{pred}}
= \rho_{r,i\to k}(\Delta)\,\widehat{\mathbf{H}}_{r,i\to k}, \\
& \mathbf{G}_i^{\text{pred}}
= \widehat{\mathbf{G}}_i,
\qquad \text{(quasi-static BS$\to$RIS LoS).}
\end{align}

Given a candidate RIS configuration $\{u_i\}$ and $\mathcal{S}_k$, the BS forms the predicted effective channel
\begin{equation}
\label{eq:Heff_predicted_short}
\bar{\mathbf{H}}_{k,\mathrm{eff}}
=
\mathbf{H}_{d,k}^{\text{pred}}
+
\sum_{i\in\mathcal{S}_k}
\mathbf{H}_{r,i\to k}^{\text{pred}}\,
\mathrm{diag}(u_i)\,
\widehat{\mathbf{G}}_i.
\end{equation}
The residual mismatch is modeled as
\begin{equation}
\label{eq:Heff_aged_short}
\mathbf{H}_{k,\mathrm{eff}}(t_0+\Delta)
=
\bar{\mathbf{H}}_{k,\mathrm{eff}}
+
\widetilde{\mathbf{E}}_k,
\quad
\mathrm{vec}(\widetilde{\mathbf{E}}_k)
\sim
\mathcal{CN}(\mathbf{0},\widetilde{\mathbf{\Sigma}}_k),
\end{equation}
where $\widetilde{\mathbf{\Sigma}}_k$ captures estimation error, CSI feedback distortion, and mobility-induced innovation.
The residual term $\widetilde{\mathbf{E}}_k$ follows the standard Gauss--Markov channel aging model, where mobility and delay-induced phase distortion are modeled as additive complex Gaussian noise \cite{TRUONG2013AGING}.

\subsection{Problem 1: Blockage Detection}
\label{subsec:prob1}

Per-RIS blockage detection at UE-$k$ uses the energy test
\begin{equation}
|Z_{k,i}|^2\ \underset{\mathcal{H}_B}{\overset{\mathcal{H}_U}{\gtrless}}\ \tau_i,
\qquad 
\widehat{\mathcal{M}}_k=\{\,i:\ |Z_{k,i}|^2\ge \tau_i\,\}.
\end{equation}
The UE reports the detected active set to the BS as a bitmap over the uplink control channel.
The total sensing-phase duration is
\begin{equation}
\label{eq:Tsense_def}
T_{\mathrm{sense}}(\ell) 
= (\ell+L_{\mathrm{cp}})T_s + T_{\mathrm{proc}} + T_{\mathrm{ctrl}}.
\end{equation}
where $T_s$ denotes the baseband sampling period, $L_{\mathrm{cp}}$ is the cyclic prefix length in samples, $T_{\mathrm{proc}}$ accounts for the UE-side processing latency required for blockage detection, and $T_{\mathrm{ctrl}}$ represents the uplink control signaling latency for reporting the detected active RIS set (bitmap) to the BS.

Thresholds are designed via an NP criterion under false-alarm constraints per panel.
The detection problem is
\begin{align}
\label{eq:prob1_detection}
\max_{\ \ell,\ \boldsymbol{\tau}}\quad & \sum_{k=1}^K \mathbb{E}\!\left[|\widehat{\mathcal{M}}_k\cap \mathcal{M}_k|\right] \\
\text{s.t.}\quad & M\le \ell,\ \ \ell\in\{2^p-1\},\ \ \text{timing constraint \eqref{eq:timing_const}}. \nonumber
\end{align}

\subsection{Problem 2: Throughput Maximization within CSI Update Interval}
\label{subsec:prob2}

\textbf{Timing constraint.}
Let $T_{\mathrm{opt}}$ denote the BS/RIS optimization runtime and $T_{\mathrm{CSI}}$ the CSI update interval set by the NR system.
The detect--optimize--transmit cycle must satisfy
\begin{equation}
\label{eq:timing_const}
T_{\mathrm{sense}}(\ell)+T_{\mathrm{opt}} < T_{\mathrm{CSI}}.
\end{equation}

\textbf{Data interval and overhead.}
The data-phase duration within each CSI update interval is given by
\begin{equation}
\label{eq:Tdata}
T_{\mathrm{data}} = T_{\mathrm{CSI}} - T_{\mathrm{sense}}(\ell) - T_{\mathrm{opt}},
\end{equation}
and the corresponding usable-data fraction is
\begin{equation}
\label{eq:gamma_def}
\gamma(\ell)
= 1 - \frac{T_{\mathrm{sense}}(\ell)+T_{\mathrm{opt}}}{T_{\mathrm{CSI}}}.
\end{equation}

\textbf{Objective (cycle-averaged sum-rate).}
Let $\mathbf{F}=[\mathbf{f}_1,\dots,\mathbf{f}_K]$ with $\|\mathbf{F}\|_F^2\le P$.
Define
\begin{equation}
\label{eq:Rk_def}
\mathbf{C}_k
=
\sigma^2\mathbf{I}
+
\sum_{j\neq k}
\mathbf{H}_{k,\mathrm{eff}}(t_0+\Delta)\mathbf{f}_j\mathbf{f}_j^H
\mathbf{H}_{k,\mathrm{eff}}^H(t_0+\Delta),
\end{equation}
and the instantaneous achievable rate
\begin{equation}
\label{eq:rate_def}
r_k
=
\log\det\!\big(\mathbf{I}+\mathbf{C}_k^{-1}
\mathbf{H}_{k,\mathrm{eff}}(t_0+\Delta)\mathbf{f}_k
\mathbf{f}_k^H\mathbf{H}_{k,\mathrm{eff}}^H(t_0+\Delta)\big).
\end{equation}

The design problem is formulated as
\begin{align}
\label{eq:prob2_sumrate}
\max_{\ \{u_i\},\,\mathbf{F}}\quad 
& \gamma(\ell)\sum_{k=1}^K
\mathbb{E}\!\left[ r_k \right] \\
\text{s.t.}\quad 
& |[u_i]_m|=1,\ \forall i,m,\quad \|\mathbf{F}\|_F^2\le P, \nonumber\\
& \mathrm{vec}(\widetilde{\mathbf{E}}_k)\sim\mathcal{CN}(\mathbf{0},\widetilde{\mathbf{\Sigma}}_k). \nonumber
\end{align}
The expectation is over the residual CSIT uncertainty in \eqref{eq:Heff_aged_short}.

\begin{figure}
\centering
    \includegraphics[width=0.99\linewidth]{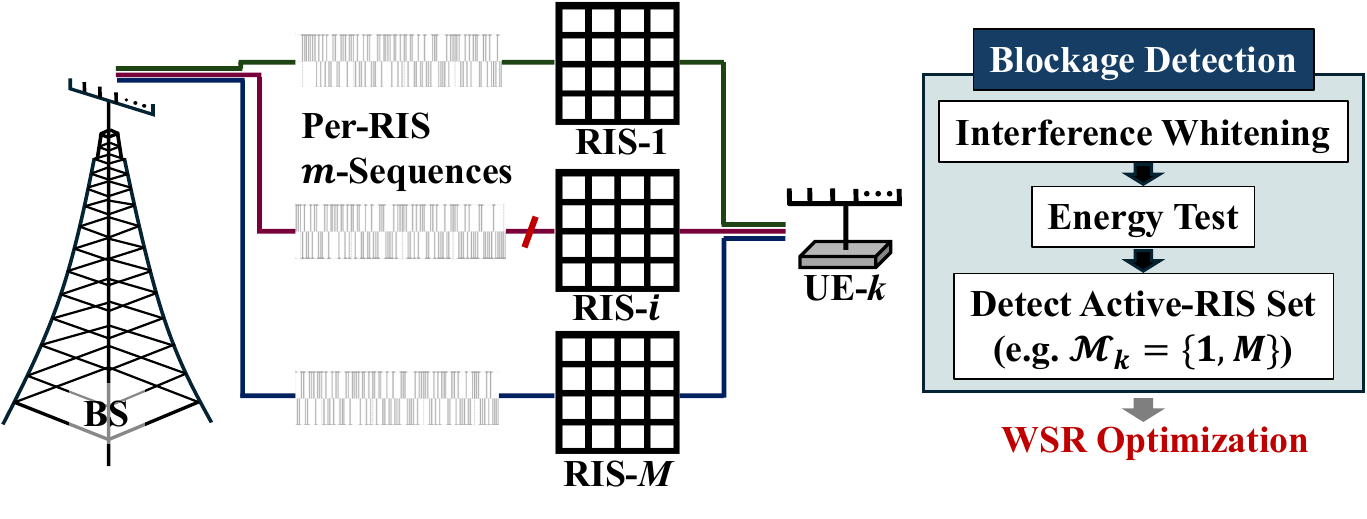}
    \caption{Per-RIS synchronization using distinct $m$-sequences, where each $s_i$ is precoded by $\mathbf v_i$ toward RIS-$i$ and each UE-$k$ detects its unblocked set $\mathcal{M}_k$ via energy detection.}
    \label{fig2}
    \vspace{-10pt}
\end{figure}

\begin{figure*}[h!t]
\captionsetup[subfloat]{farskip=2pt}
\vspace{-7pt}
\centering
        \subfloat[$M=8$, $K=4$, $\ell=127$]{\includegraphics[width=.33\linewidth]{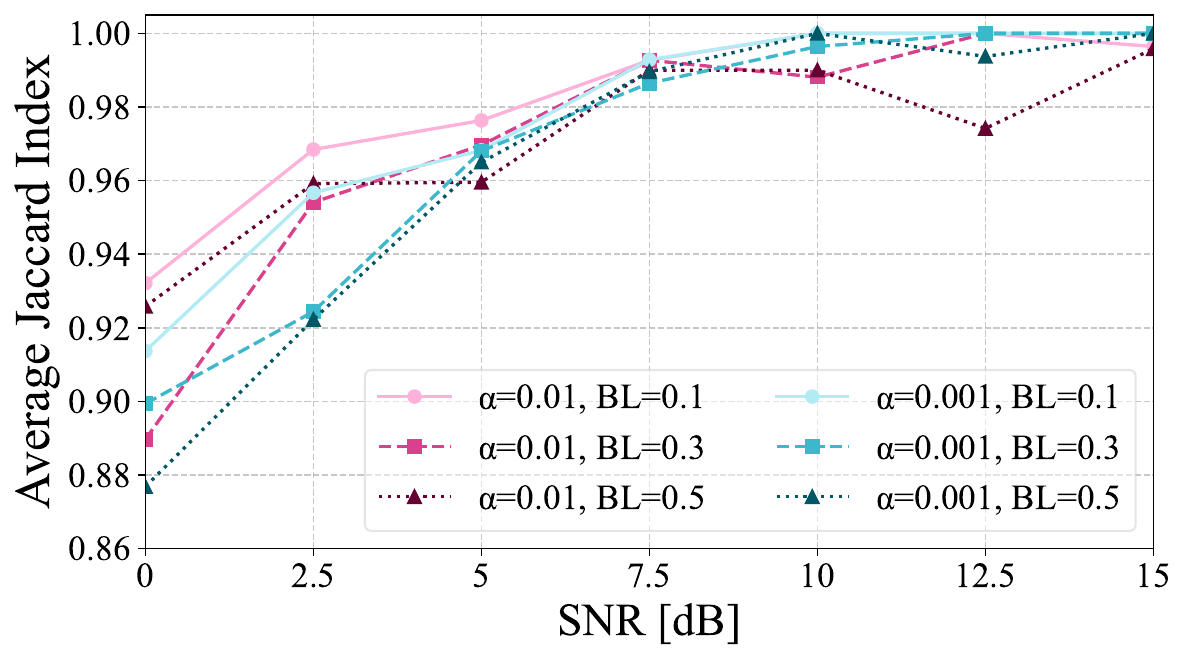}}
        \subfloat[$M=16$, $K=8$, $\ell=127$]{\includegraphics[width=.33\linewidth]{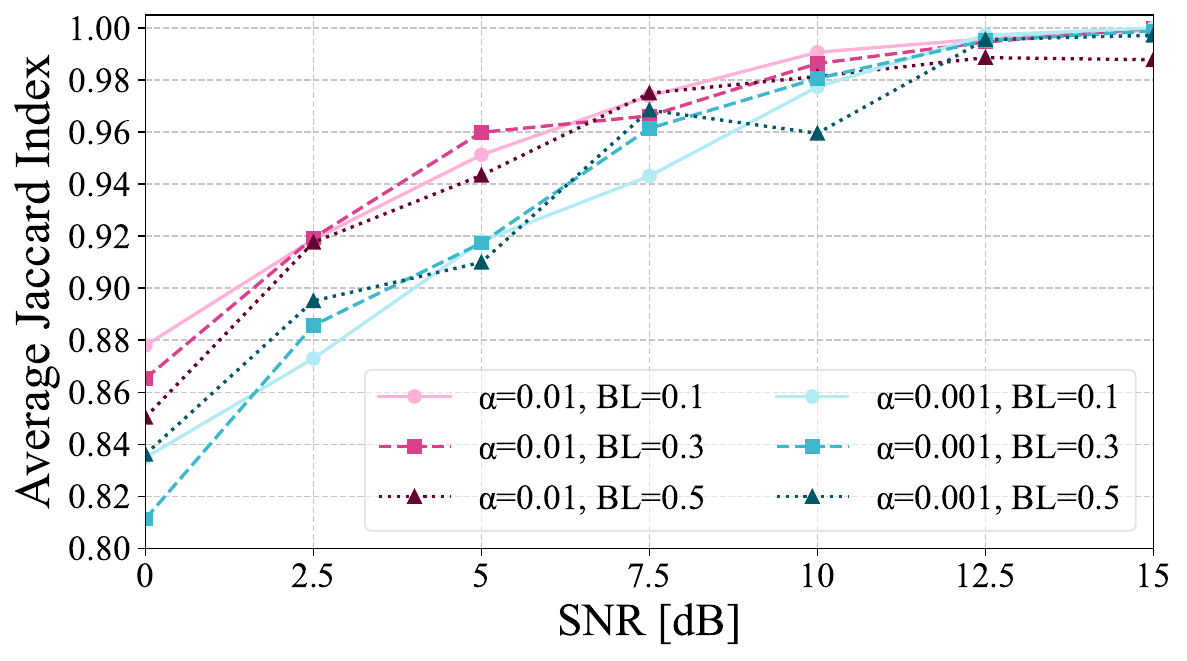}}
        \subfloat[$M=32$, $K=16$, $\ell=127$]{\includegraphics[width=.33\linewidth]{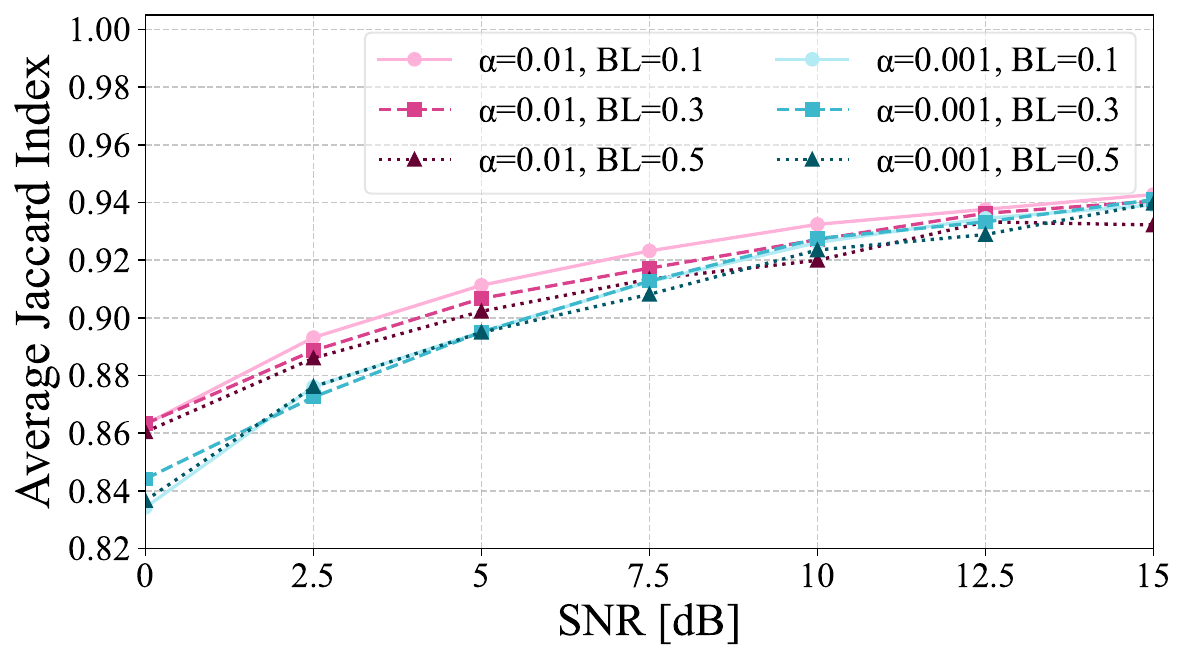}}\\
        \subfloat[$M=8$, $K=4$, $\ell=255$]{\includegraphics[width=.33\linewidth]{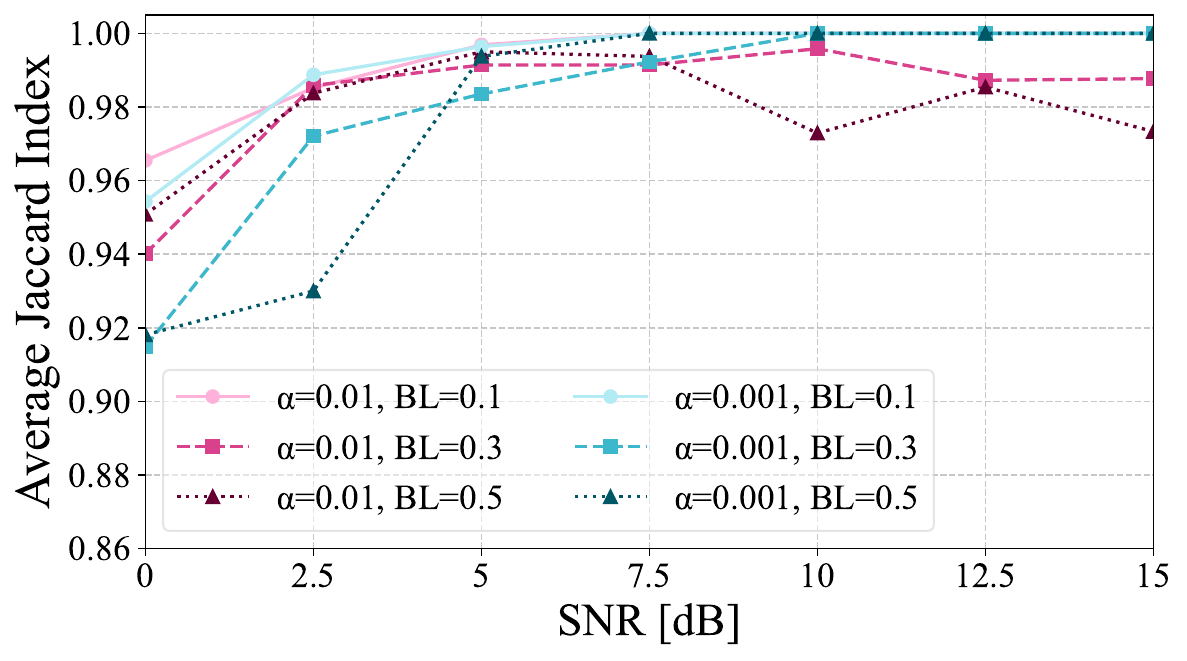}}
        \subfloat[$M=16$, $K=8$, $\ell=255$]{\includegraphics[width=.33\linewidth]{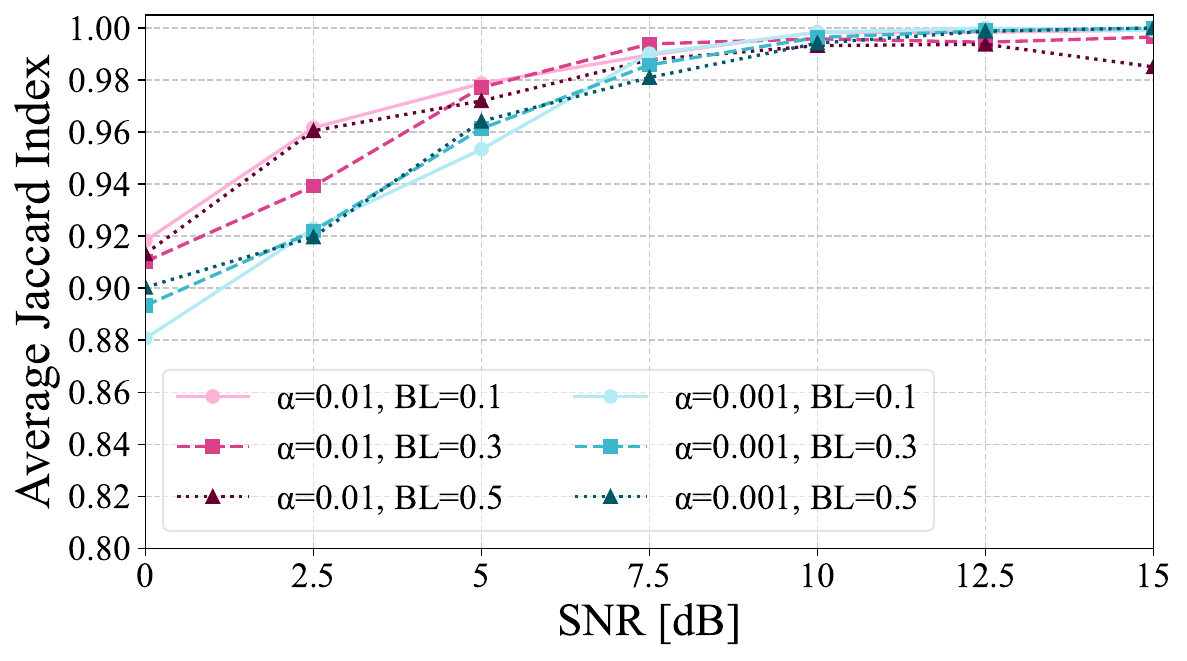}}
        \subfloat[$M=32$, $K=16$, $\ell=255$]{\includegraphics[width=.33\linewidth]{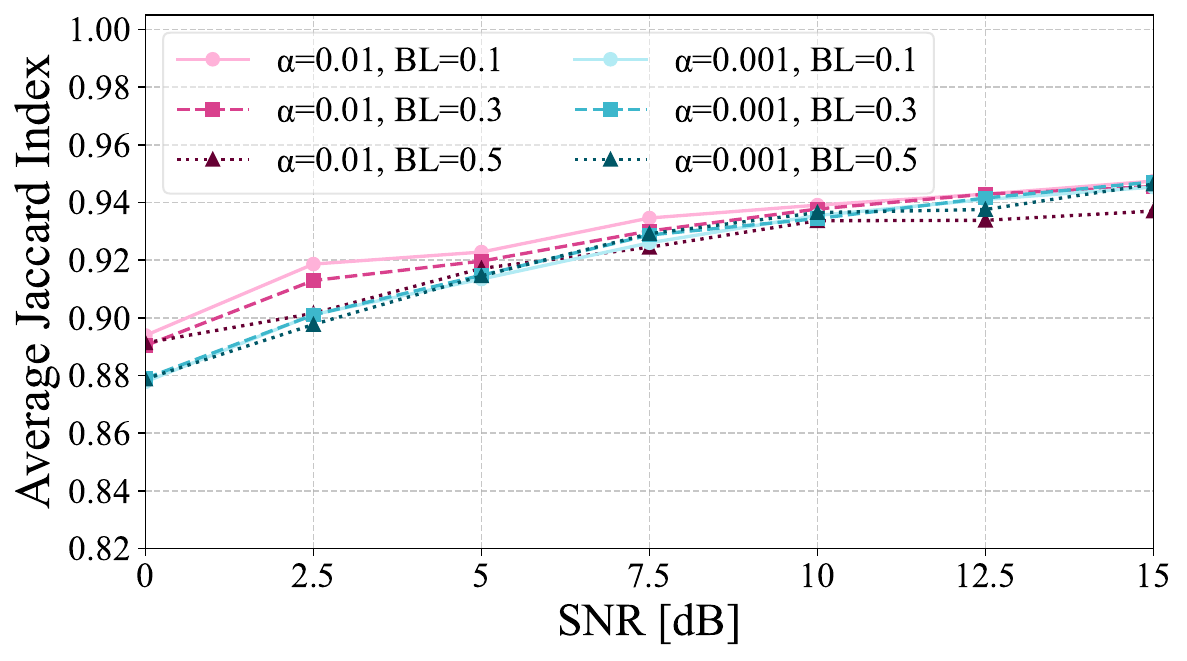}}
\vspace{-3pt}
\caption{Blockage-detection performance measured by the Jaccard index \cite{JACCARD1901} versus SNR for different numbers of UEs ($K$), RIS panels ($M$), and sequence lengths $\ell$, under the target false-alarm probability $\alpha$ and per-RIS blockage probability (BL).}
\vspace{-2pt}
\label{fig:SNR_BDR}
\end{figure*}

\begin{figure}[h!t]
\captionsetup[subfloat]{farskip=2pt}
\vspace{-7pt}
\centering
        \subfloat[$M=8$, $K=4$]{\includegraphics[width=.7\linewidth]{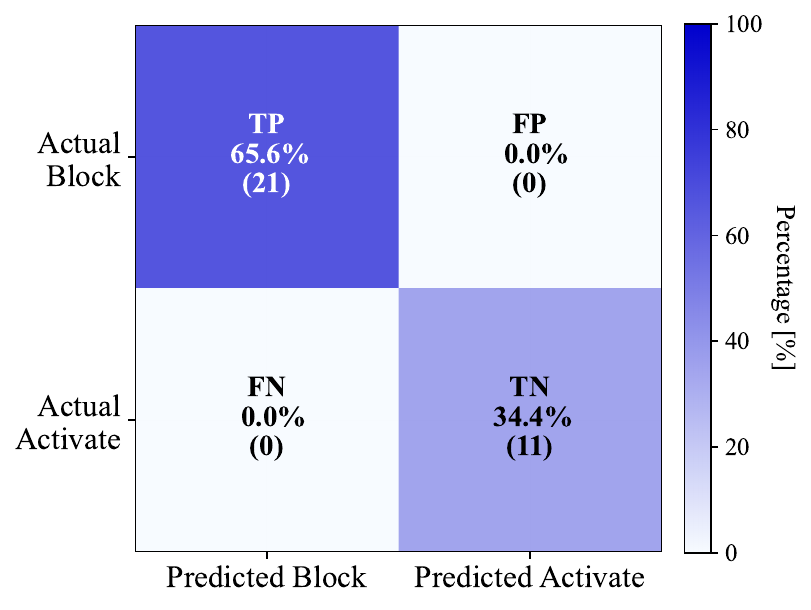}}\\
        \subfloat[$M=16$, $K=8$]{\includegraphics[width=.7\linewidth]{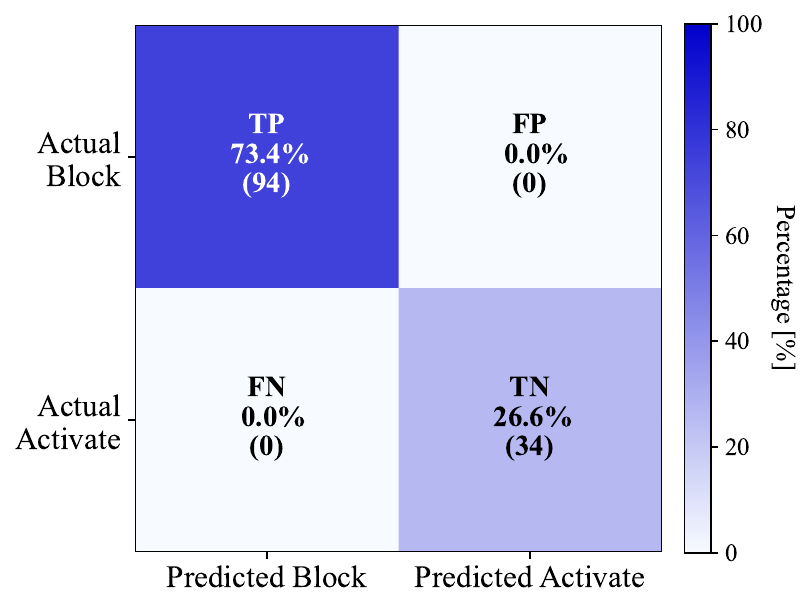}}\\
        \subfloat[$M=32$, $K=16$]{\includegraphics[width=.7\linewidth]{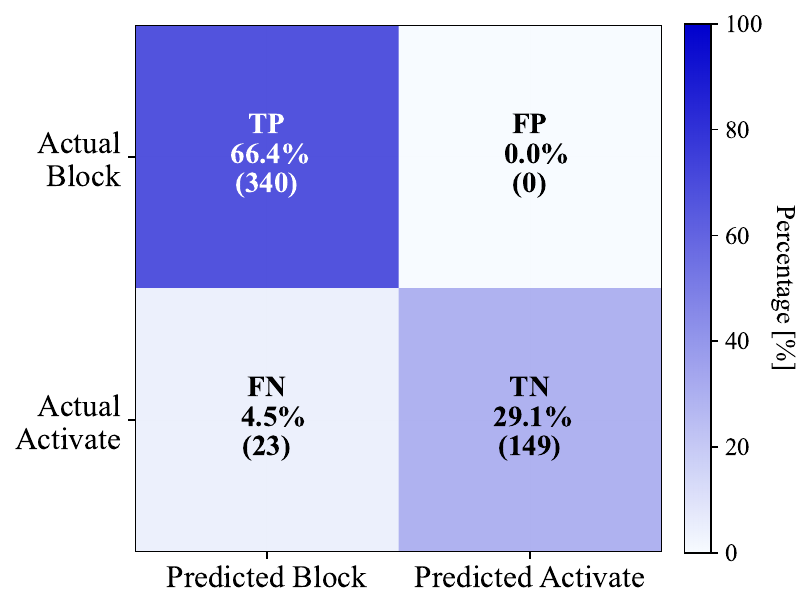}}
\vspace{-3pt}
\caption{Visualization of RIS activation/blocking decisions for all UEs under $\alpha = 10^{-3}$ and $\ell = 255$, illustrating true/false positives and negatives.}
\vspace{-2pt}
\label{fig:BD_TFPN}
\end{figure}

\section{Blockage Detection Framework}
\label{sec:blockage_detection}

\subsection{RIS-Indexed Synchronization Sequences}
\label{subsec:indexed_sequences}

Transmission employs short CP-assisted indexed synchronization sequences to identify the active (unblocked) RIS panels with low sensing overhead.
All sensing correlations are computed \emph{after CP removal} over the $\ell$ useful samples, so that the effective multipath convolution is circular and to preserve the periodic correlation property of cyclic shifts, as in Fig.~\ref{fig2}.

\paragraph*{Timing and CP assumption}
The model assumes that, during the sensing phase, the propagation delays of all relevant paths---including the direct BS$\!\to$UE link and the cascaded BS$\!\to$RIS$\!\to$UE links associated with candidate RIS panels---are contained within the cyclic prefix of length $L_{\mathrm{cp}}$.
This assumption can be ensured in practice by choosing $L_{\mathrm{cp}}$ to exceed the maximum expected delay spread given the known BS and RIS locations.
Under this condition, CP removal converts linear convolution into circular convolution over the useful symbol interval, and the cyclic-shift correlation structure of the sequences is exactly preserved.
Any residual energy from paths exceeding the CP duration can be absorbed
into an effective colored disturbance, and is neglected for clarity.

\textbf{$m$-sequence construction and indexing.}
Let $\ell=2^p-1$ and let $\{x[n]\}$ be a binary sequence generated by a
primitive degree-$p$ LFSR:
\begin{equation}
x[n+p] = \bigoplus_{r \in \mathcal{F}} x[n+r],
\end{equation}
where $\mathcal{F}$ denotes the feedback-tap set and $\oplus$ denotes
modulo-2 addition.
Define the BPSK $m$-sequence $s[n]=(-1)^{x[n]}$, which satisfies the periodic
correlation property
\begin{equation}
\sum_{n=0}^{\ell-1} s[n]\, s^{\!*}[(n+\Delta)\bmod\ell]
=
\begin{cases}
\ell, & \Delta = 0,\\
-1, & \Delta \neq 0.
\end{cases}
\label{eq:mseq_corr}
\end{equation}
RIS-$i$ is tagged by a cyclic shift $\nu_i=i-1$,
\begin{equation}
s_i[n] = s[(n+\nu_i)\bmod \ell],
\qquad i=1,\dots,M,
\qquad M\le \ell.
\label{eq:shifted_mseq}
\end{equation}
The constant cross-correlation bias ($-1/\ell$) enables index separation but
induces a deterministic coupling among indices; this coupling is explicitly
accounted for and removed in
Section~\ref{subsec:decoupled_detection}.

\textbf{Pilot transmission.}
During sensing, the BS transmits a superposition of indexed pilots
\begin{equation}
\mathbf{x}[n]
=
\sum_{i=1}^{M} \mathbf{v}_i\, s_i[n],
\qquad
\sum_{i=1}^{M}\|\mathbf{v}_i\|_2^2 \le P_{\mathrm{pilot}},
\label{eq:pilot_superposition}
\end{equation}
where $\mathbf{v}_i\in\mathbb{C}^{N_t}$ is a predefined probing/steering vector toward RIS-$i$, computable offline from the known RIS locations, as in Fig.~\ref{fig2}.
Each RIS employs a fixed sensing configuration (e.g., all-pass reflection), so that the per-RIS effective sensing channel remains constant over the sensing interval.

\subsection{Matched Filtering Model and NP Detection Metric}
\label{subsec:np_threshold}

\textbf{Received signal after CP removal.}
The received sensing signal at UE-$k$ is
\begin{equation}
\mathbf{y}_k[n]
=
\sum_{i=1}^{M}
\mathbf{H}^{\mathrm{eff}}_{k,i}\mathbf{v}_i\, s_i[n]
+
\mathbf{n}_k[n],
\qquad
\mathbf{n}_k[n]\sim\mathcal{CN}(\mathbf{0},\sigma^2\mathbf{I}),
\label{eq:rx_pilot}
\end{equation}
where $\mathbf{H}^{\mathrm{eff}}_{k,i}$ denotes the effective BS--RIS-$i$--UE
channel under the sensing configuration.
UE-$k$ applies a fixed linear combiner, which uses equal-gain combining
\begin{equation}
\tilde{y}_k[n]
=
\tfrac{1}{\sqrt{N_{r,k}}}\mathbf{1}^{H}\mathbf{y}_k[n].
\label{eq:egc}
\end{equation}

\textbf{Matched-filter outputs and index coupling.}
Define
\begin{equation}
a_{k,i}
\triangleq
\tfrac{1}{\sqrt{N_{r,k}}}
\mathbf{1}^{H}\mathbf{H}^{\mathrm{eff}}_{k,i}\mathbf{v}_i,
\label{eq:a_def}
\end{equation}
and
\begin{equation}
Z_{k,i}
\triangleq
\sum_{n=0}^{\ell-1} s_i^{\!*}[n]\tilde{y}_k[n].
\label{eq:Z_def}
\end{equation}
Stacking $\mathbf{Z}_k=[Z_{k,1},\dots,Z_{k,M}]^T$ and
$\mathbf{a}_k=[a_{k,1},\dots,a_{k,M}]^T$ yields
\begin{equation}
\mathbf{Z}_k
=
\mathbf{C}\mathbf{a}_k + \tilde{\mathbf{n}}_k,
\qquad
\tilde{\mathbf{n}}_k\sim\mathcal{CN}(\mathbf{0},\sigma^2\mathbf{C}),
\label{eq:Z_vector_model}
\end{equation}
with
\begin{equation}
\mathbf{C}=(\ell{+}1)\mathbf{I}_M-\mathbf{J}_M,
\label{eq:C_def}
\end{equation}
where $\mathbf{J}_M$ is the all-ones matrix.
This expression makes explicit the deterministic inter-index coupling
induced by the constant cross-correlation of $m$-sequences.

\subsection{Decoupled Detection via Inversion of the Coupling Matrix}
\label{subsec:decoupled_detection}

Since $\mathbf{C}$ is invertible for $M<\ell{+}1$, the deterministic coupling vanishes via
\begin{equation}
\widehat{\mathbf{a}}_k
\triangleq
\mathbf{C}^{-1}\mathbf{Z}_k
=
\mathbf{a}_k+\mathbf{C}^{-1}\tilde{\mathbf{n}}_k,
\label{eq:ahat_def}
\end{equation}
which yields unbiased per-panel estimates.
Each component satisfies
$\widehat{a}_{k,i}\sim\mathcal{CN}(a_{k,i},\sigma^2[\mathbf{C}^{-1}]_{ii})$,
allowing a per-panel NP energy test
\begin{equation}
|\widehat{a}_{k,i}|^2
\ \underset{\mathcal{H}_{B,i}}{\overset{\mathcal{H}_{U,i}}{\gtrless}}\
\tau_i,
\qquad
\tau_i=-\sigma^2[\mathbf{C}^{-1}]_{ii}\ln\alpha_i,
\label{eq:NP_tau}
\end{equation}
which determines the active RIS set for UE-$k$ as in Fig.~\ref{fig2}.

\subsection{Sequence-Length and Latency Trade-offs}
\label{subsec:length_latency}

From \eqref{eq:Z_def}, the post-correlation SNR scales approximately linearly
with $\ell$ in the noise-limited regime.
A sensing transmission occupies airtime
$T_{\mathrm{air}}(\ell)=(\ell+L_{\mathrm{cp}})T_s$, and feasibility requires
\begin{equation}
\label{eq:time_constraints}
T_{\mathrm{sense}}(\ell)+T_{\mathrm{opt}}<T_{\mathrm{CSI}},
\qquad
M<\ell{+}1.
\end{equation}
Here, $T_{\mathrm{CSI}}$ denotes the CSI update interval configured by the NR system.
A standard-aligned choice is $\ell=127$, consistent with the NR PSS sequence length \cite{TS138211};
longer sequences can be employed when larger $M$ or stricter blockage-detection reliability is required, provided that the resulting sensing and optimization latency remains within the CSI update interval.

\subsection{Empirical Evaluation of Blockage Detection}

All blockage-detection results are evaluated using geometry-consistent
mmWave channels with user mobility and RIS link blocking, generated by a
ray-tracing-based simulator described in Section~\ref{subsec:dataset}.

Fig.~\ref{fig:SNR_BDR} shows that the proposed detector achieves high accuracy even with short sequences, and the performance improves with SNR and sequence length.
The accuracy further increases for smaller numbers of RIS panels and UEs due to the reduced multi-user and inter-RIS interference.

Fig.~\ref{fig:BD_TFPN} reports the true and false-positive detection performance with the threshold parameter set to $\alpha=10^{-3}$.
Under this setting, the proposed detector successfully identifies the active RIS panels with a probability exceeding $95\%$, while incurring virtually no false positives across all tested configurations.
Suppressing false positives is particularly important in RIS-assisted systems, since transmitting toward inactive or blocked RIS panels may result in severe signal attenuation and wasted transmission resources.
These results demonstrate that the proposed approach provides scalable and highly reliable active RIS detection under realistic mobility and
blockage conditions.

\section{SCRPA: Robust RIS--Precoder Alternation Under Aged CSIT}
\label{sec:scrpa}

This section addresses the cycle-averaged \emph{sum-rate} maximization problem in \eqref{eq:prob2_sumrate}, which maximization jointly optimizes the BS precoder $\mathbf{F}$ and the RIS phase shifts $\{u_i\}$ under
(i) per-element unit-modulus RIS constraints,
(ii) multi-RIS structured effective channels in \eqref{eq:Heff_compact},
and (iii) temporally aged and imperfect CSIT modeled by the Gaussian residual uncertainty in \eqref{eq:Heff_aged_short}.
The objective is to maximize the robust cycle-averaged throughput under time-varying CSI uncertainty induced by feedback delay and user mobility.

This optimization introduces a stochastic block MM procedure termed \emph{stochastic coordinated RIS--precoder alternation} (SCRPA).
SCRPA adopts a sample-average approximation (SAA) of the stochastic objective and applies block-coordinate updates that admit closed-form steps for both the BS precoder and the RIS phases.
An optional $\beta$-bit projection step is further included to support practical finite-resolution RIS hardware.

\vspace{4pt}
\noindent\textbf{Monte Carlo ensemble for robust design.}
The expectation over the random effective channels in \eqref{eq:prob2_sumrate} is generally intractable in closed form.
SCRPA therefore uses an SAA to explicitly optimize the expected sum-rate under stochastic CSIT uncertainty.
At SCRPA iteration $t$, given the current RIS phases $\{u_i^{(t)}\}$, the BS constructs the aged channel prediction
\begin{equation}
\label{eq:Heff_bar_iter}
\bar{\mathbf{H}}_{k,\mathrm{eff}}^{(t)}
=
\mathbf{H}_{d,k}^{\text{pred}}
+
\sum_{i\in\widehat{\mathcal{M}}_k}
\mathbf{H}_{r,i\to k}^{\text{pred}}\,
\mathrm{diag}(u_i^{(t)})\,
\widehat{\mathbf{G}}_i,
\end{equation}
and draws $S$ i.i.d.\ channel realizations
\begin{equation}
\label{eq:Hkstochastic_scrpa}
\mathbf{H}_{k,\mathrm{eff}}^{(s,t)}
=
\bar{\mathbf{H}}_{k,\mathrm{eff}}^{(t)}
+
\widetilde{\mathbf{E}}_k^{(s)},
\qquad
\widetilde{\mathbf{E}}_k^{(s)} \sim \mathcal{CN}(\mathbf{0},\widetilde{\mathbf{\Sigma}}_k),
\end{equation}
for $s=1,\dots,S$.
Larger $S$ improves SAA fidelity to the underlying expectation, with computational complexity scaling approximately linearly in $S$.
In practice, moderate budgets (e.g., $S{=}4$--$16$) are sufficient to stabilize performance while remaining compatible with the beam-coherence-time constraint in \eqref{eq:timing_const}.

For a given precoder $\mathbf{F}=[\mathbf{f}_1,\dots,\mathbf{f}_K]$, define,
for UE~$k$ and sample $s$,
\begin{equation}
\label{eq:qR_def_scrpa}
\mathbf{q}_k^{(s,t)}
\triangleq
\mathbf{H}_{k,\mathrm{eff}}^{(s,t)} \mathbf{f}_k,
\qquad
\mathbf{C}_k^{(s,t)}
=
\sigma^2\mathbf{I}
+
\sum_{j\neq k}
\mathbf{H}_{k,\mathrm{eff}}^{(s,t)}\mathbf{f}_j
\mathbf{f}_j^H
{\mathbf{H}_{k,\mathrm{eff}}^{(s,t)}}^{\!H}.
\end{equation}
The sampled instantaneous rate is
\begin{equation}
\label{eq:rate_sampled}
r_k^{(s,t)}
=
\log\det\!\Big(
\mathbf{I}
+
\big(\mathbf{C}_k^{(s,t)}\big)^{-1}
\mathbf{q}_k^{(s,t)} {\mathbf{q}_k^{(s,t)}}^{\!H}
\Big).
\end{equation}
Accordingly, \eqref{eq:prob2_sumrate} is approximated by the SAA objective
(up to the known scaling factor $\gamma(\ell)$):
\begin{equation}
\label{eq:SAA_sumrate_obj}
\max_{\{u_i\},\,\mathbf{F}}
\quad
\frac{1}{S}\sum_{s=1}^S \sum_{k=1}^K r_k^{(s,t)}.
\end{equation}

\vspace{4pt}
\noindent\textbf{WMMSE reformulation under stochastic CSIT.}
For each sample $s$, the classical WSR/WMMSE equivalence \cite{SHI2011}, 
which builds on the fundamental rate–MSE relationship underlying GDFE \cite{CIOFFI1997GDFE}, 
yields an equivalent weighted-MSE surrogate whose stationary points coincide with those of the sampled sum-rate problem.
Introducing for each UE~$k$ and sample $s$ an MMSE receive filter
$\mathbf{U}_k^{(s,t)}$ and a weight matrix $\mathbf{W}_k^{(s,t)}\succeq\mathbf{0}$,
define the MSE matrix
\begin{align}
\label{eq:MSE_scrpa}
\mathbf{E}_k^{(s,t)}
&=
\mathbf{I}
-
{\mathbf{U}_k^{(s,t)}}^{\!H}
\mathbf{H}_{k,\mathrm{eff}}^{(s,t)}\mathbf{f}_k
-
\mathbf{f}_k^{\!H}
{\mathbf{H}_{k,\mathrm{eff}}^{(s,t)}}^{\!H}
\mathbf{U}_k^{(s,t)} \nonumber\\
&\quad
+
{\mathbf{U}_k^{(s,t)}}^{\!H}
\Big(
\mathbf{H}_{k,\mathrm{eff}}^{(s,t)}\mathbf{F}\mathbf{F}^{\!H}
{\mathbf{H}_{k,\mathrm{eff}}^{(s,t)}}^{\!H}
+
\sigma^2\mathbf{I}
\Big)
\mathbf{U}_k^{(s,t)}.
\end{align}
At iteration $t$, SCRPA minimizes the SAA WMMSE objective
\begin{equation}
\label{eq:SAA_obj_scrpa}
\widehat{f}^{(t)}
=
\frac{1}{S}
\sum_{s=1}^S
\sum_{k=1}^K
\Big(
\mathrm{Tr}\!\big(\mathbf{W}_k^{(s,t)}\mathbf{E}_k^{(s,t)}\big)
-
\log\det\mathbf{W}_k^{(s,t)}
\Big),
\end{equation}
which is a tight surrogate of the negative SAA sum-rate and enables
closed-form block updates.

\vspace{4pt}
\noindent\textbf{Block-MM iteration (SCRPA).}
SCRPA applies block coordinate descent / block MM to iteratively decrease \eqref{eq:SAA_obj_scrpa}:

\textit{Channel sampling:}
Given $\{u_i^{(t)}\}$, construct $\bar{\mathbf{H}}_{k,\mathrm{eff}}^{(t)}$ via \eqref{eq:Heff_bar_iter} and draw $\{\mathbf{H}_{k,\mathrm{eff}}^{(s,t)}\}_{s=1}^S$ according to \eqref{eq:Hkstochastic_scrpa}.

\textit{(A) Receiver/weight update:}
For each $k$ and $s$,
\begin{align}
\label{eq:update_u}
& \mathbf{U}_k^{(s,t)}
\leftarrow
\Big(
\mathbf{H}_{k,\mathrm{eff}}^{(s,t)}\mathbf{F}^{(t)}
(\mathbf{H}_{k,\mathrm{eff}}^{(s,t)}\mathbf{F}^{(t)})^{\!H}
+
\sigma^2\mathbf{I}
\Big)^{-1}
\mathbf{H}_{k,\mathrm{eff}}^{(s,t)}\mathbf{f}_k^{(t)},\\
\label{eq:update_w}
& \mathbf{W}_k^{(s,t)}
\leftarrow
\big(\mathbf{E}_k^{(s,t)}\big)^{-1}.
\end{align}

\textit{(B) Precoder update:}
Define
\begin{align}
\label{eq:AB_def_nofair}
& \mathbf{A}^{(t)}
=
\sum_{k=1}^K
\sum_{s=1}^S
{\mathbf{H}_{k,\mathrm{eff}}^{(s,t)}}^{\!H}
\mathbf{U}_k^{(s,t)}\mathbf{W}_k^{(s,t)}
{\mathbf{U}_k^{(s,t)}}^{\!H}
\mathbf{H}_{k,\mathrm{eff}}^{(s,t)}, \\
& \mathbf{B}^{(t)}
=
\big[
\mathbf{b}_1^{(t)},\dots,\mathbf{b}_K^{(t)}
\big], \qquad
\mathbf{b}_k^{(t)}
\triangleq
\sum_{s=1}^S
{\mathbf{H}_{k,\mathrm{eff}}^{(s,t)}}^{\!H}
\mathbf{U}_k^{(s,t)}\mathbf{W}_k^{(s,t)} .
\end{align}
Then the BS precoder is updated as
\begin{equation}
\label{eq:update_F_nofair}
\mathbf{F}^{(t+1)}
=
\big(\mathbf{A}^{(t)}+\lambda\mathbf{I}\big)^{-1}\mathbf{B}^{(t)},
\qquad
\|\mathbf{F}^{(t+1)}\|_F^2=P,
\end{equation}
where $\lambda\ge 0$ is chosen (e.g., by bisection) to satisfy the power
constraint.

\textit{(C) RIS phase update (continuous or quantized):}
Fix a panel index $i$ and all other blocks at their current values.
Since $\widehat{f}^{(t)}$ is generally nonconvex in $u_i$, SCRPA applies an MM
step based on a quadratic majorizer. Let $g_i^{(t)}$ denote the (sample-average)
gradient estimate with respect to $u_i$, defined in \eqref{eq:gradient}.
Assuming that $\nabla_{u_i}\widehat{f}^{(t)}$ is $L_i$-Lipschitz (or enforcing it
via backtracking on $L_i$), the following quadratic surrogate majorizes
$\widehat{f}^{(t)}(u_i)$:
\begin{equation}
\label{eq:MM_majorizer}
\begin{aligned}
\widehat{f}^{(t)}(u_i)
\le\;
\widehat{f}^{(t)}(u_i^{(t)})
&+
\mathrm{Re}\!\left\{
\left(g_i^{(t)}\right)^{\!H}
\left(u_i-u_i^{(t)}\right)
\right\} \\
&+
\frac{L_i}{2}
\left\|u_i-u_i^{(t)}\right\|_2^2 .
\end{aligned}
\end{equation}
Minimizing \eqref{eq:MM_majorizer} over the unit-modulus set yields the closed-form continuous-phase update
\begin{equation}
\label{eq:update_closed}
[u_i^{(t+1)}]_m^{\mathrm{cont}}
=
\exp\!\big(j\arg([g_i^{(t)}+L_i u_i^{(t)}]_m)\big).
\end{equation}
Practical RIS hardware support optionally projects the continuous update onto the $\beta$-bit alphabet $\mathcal{U}_\beta$ elementwise.
While the continuous update guarantees monotonic descent of $\widehat{f}^{(t)}$, the projection may slightly violate strict monotonicity;
nevertheless, backtracking on $L_i$ yields stable behavior in practice.

\vspace{2pt}
\noindent\textbf{Stochastic gradient used in (C).}
For completeness, the stochastic gradient estimate for RIS panel $i$ is
\begin{equation}
\label{eq:gradient}
\begin{aligned}
g_i^{(t)}
&=
\frac{1}{S}
\sum_{s=1}^{S}
\sum_{k\in\mathcal{K}_i}
\mathrm{diag}\!\Big(
{\mathbf{H}_{r,i\to k}^{\mathrm{pred}}}^{\!H}
\mathbf{Z}_k^{(s,t)}
\widehat{\mathbf{G}}_i^{\!H}
\Big),
\end{aligned}
\end{equation}
with
\begin{equation}
\label{eq:Z_def}
\mathbf{Z}_k^{(s,t)}
\triangleq
\mathbf{C}_k^{(s,t)}
-
\boldsymbol{\Psi}_k^{(s,t)}
\mathbf{H}_{k,\mathrm{eff}}^{(s,t)}
\mathbf{Q}^{(t)} ,
\end{equation}
where
$\boldsymbol{\Psi}_k^{(s,t)}
=
\mathbf{U}_k^{(s,t)}\mathbf{W}_k^{(s,t)}{\mathbf{U}_k^{(s,t)}}^{\!H}$,
$\mathbf{C}_k^{(s,t)}
=
\mathbf{U}_k^{(s,t)}\mathbf{W}_k^{(s,t)}{\mathbf{f}_k^{(t)}}^{\!H}$,
and $\mathbf{Q}^{(t)}=\mathbf{F}^{(t)}{\mathbf{F}^{(t)}}^{\!H}$.

\begin{algorithm}[t]
\caption{Quantization-Compatible SCRPA (Robust Sum-Rate Maximization)}
\label{alg:var_scrpa}
\begin{algorithmic}[1]
\State \textbf{Input:} $\{\widehat{\mathcal{M}}_k\}$,
$\{\mathbf{H}_{d,k}^{\text{pred}},\mathbf{H}_{r,i\to k}^{\text{pred}},\widehat{\mathbf{G}}_i\}$,
$\{\widetilde{\mathbf{\Sigma}}_k\}$,
$P$,
$S$,
$\beta$.
\State \textbf{Init:} feasible $\{u_i^{(0)}\}$, $\mathbf{F}^{(0)}$.
\For{$t=0,\dots,T_{\max}$}
    \State Sample stochastic channels via \eqref{eq:Heff_bar_iter}--\eqref{eq:Hkstochastic_scrpa}.
    \State Update $\{\mathbf{U}_k^{(s,t)}\}$ and $\{\mathbf{W}_k^{(s,t)}\}$ via \eqref{eq:update_u}--\eqref{eq:update_w}.
    \State Update $\mathbf{F}^{(t+1)}$ via \eqref{eq:update_F_nofair}.
    \State Update $\{u_i^{(t+1)}\}$ via \eqref{eq:update_closed} and (optionally) $\beta$-bit projection.
\EndFor
\State \textbf{Output:} $\mathbf{F}^\star$, $\{u_i^\star\}$.
\end{algorithmic}
\end{algorithm}

\vspace{4pt}
\noindent\textbf{Complexity and convergence.}
Each SCRPA iteration performs stochastic channel sampling, WMMSE-based receiver/weight updates, a BS precoder update, and RIS phase updates.

\textit{Per-iteration complexity.}
The dominant costs are
\[
\mathcal{O}\!\Big(S\sum_{k=1}^K N_{r,k}^3\Big)
\]
for the receiver updates,
\[
\mathcal{O}\!\Big(S\sum_{k=1}^K N_{r,k}N_t^2\Big) + \mathcal{O}(N_{\rm bis}N_t^3)
\]
for forming $\mathbf{A}^{(t)},\mathbf{B}^{(t)}$ and updating the precoder, and
\[
\mathcal{O}\!\Big(S\sum_{i=1}^M\sum_{k\in\mathcal{K}_i} N_i N_{r,k} N_t\Big)
\]
for the RIS phase updates, where $N_{\rm RIS}\triangleq\sum_{i=1}^M N_i$.
The optional $\beta$-bit projection adds $\mathcal{O}(N_{\rm RIS})$.

Overall, the complexity scales linearly with the Monte Carlo budget $S$ and $N_{\rm RIS}$, and cubically with $N_t$ through the precoder solve. All SAA operations are parallelizable across samples.

\textit{Convergence.}
Under continuous phases, with $L_i$ chosen so that \eqref{eq:MM_majorizer}
majorizes $\widehat{f}^{(t)}$, the block-MM updates yield a monotonically
nonincreasing sequence of objective values in \eqref{eq:SAA_obj_scrpa} and
converge to a block-stationary point of the fixed-ensemble SAA problem.
Under finite-resolution RIS constraints, each RIS-panel MM subproblem is
solved on the discrete set via the projection step, and SCRPA converges to a
discrete block-stationary point under the adopted updates.

\begin{theorem}[Convergence of SCRPA]
\label{thm:conv_scrpa}
Fix an outer iteration index $t$ and the Monte Carlo ensemble $\{\mathbf{H}_{k,\mathrm{eff}}^{(s,t)}\}_{k,s}$ generated by \eqref{eq:Hkstochastic_scrpa}. Consider SCRPA with cyclic block updates over $\{\mathbf{U}_k^{(s)},\mathbf{W}_k^{(s)}\}_{k,s}$, $\mathbf{F}$, and $\{u_i\}$.
\begin{enumerate}
\item \textbf{Continuous RIS phases.}
Under $|[u_i]_m|=1$ and with $L_i$ chosen so that \eqref{eq:MM_majorizer}
majorizes $\widehat{f}^{(t)}$ in \eqref{eq:SAA_obj_scrpa}, the iterates satisfy
\[
\widehat{f}^{(t)}(\mathcal{V}^{r+1})
\le
\widehat{f}^{(t)}(\mathcal{V}^{r}),
\]
and hence converge. Every accumulation point is block-stationary for the fixed-ensemble SAA problem.
\item \textbf{Finite-resolution RIS phases.}
Under $[u_i]_m\in\mathcal{U}_\beta$, SCRPA monotonically decreases the MM surrogate and converges to a discrete block-stationary point.
\end{enumerate}
\end{theorem}

\begin{figure}[ht]
\captionsetup[subfloat]{farskip=2pt}
\vspace{-7pt}
\centering
    \includegraphics[width=0.96\linewidth]{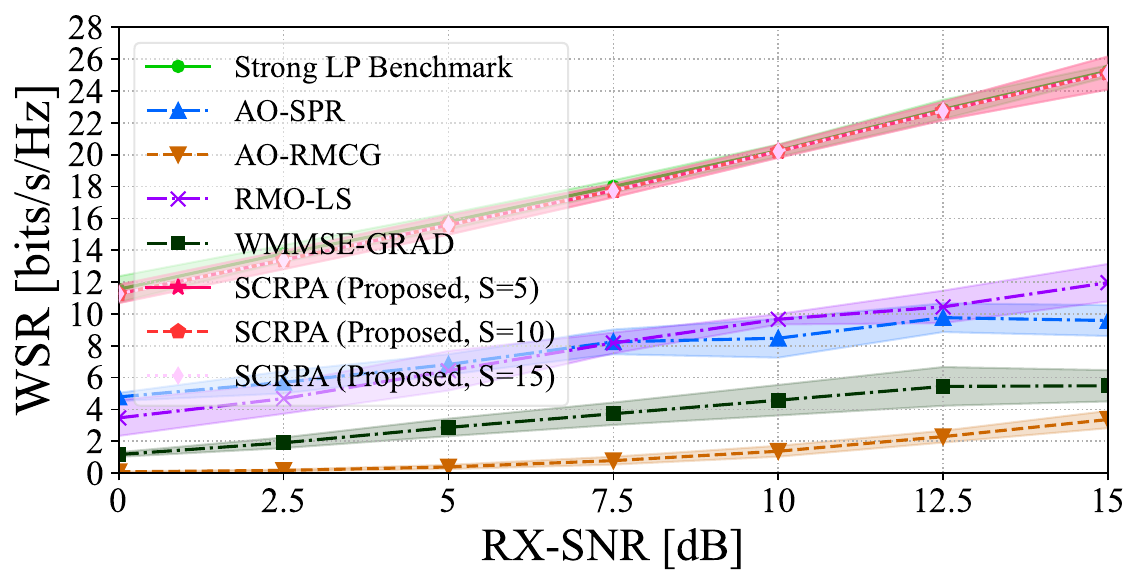}
\vspace{-3pt}
\caption{WSR versus RX-SNR for UEs moving at 5 km/h ($M=16$, $K=8$) under 4-bit RIS phase quantization, with all algorithms exploiting all RIS links. SCRPA results for $S=5$, $10$, and $15$. Markers indicate the mean over 50 channel realizations, with error bars showing one standard deviation.} 
\vspace{-2pt}
\label{fig:WSR_SNR_ALL}
\end{figure}

\begin{figure}[ht]
\captionsetup[subfloat]{farskip=2pt}
\vspace{-7pt}
\centering
    \includegraphics[width=0.89\linewidth]{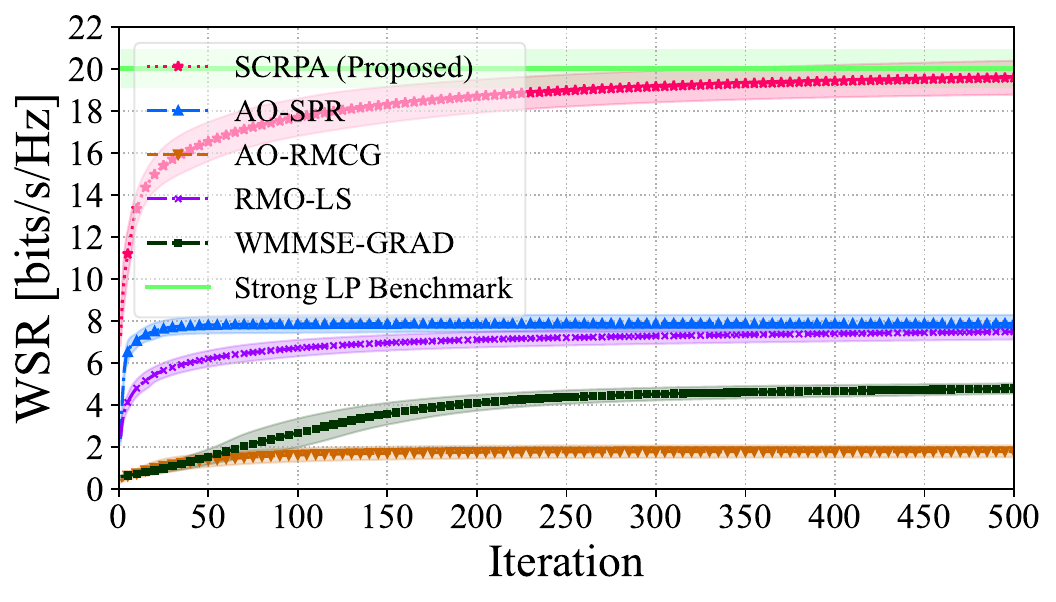}
\vspace{-3pt}
\caption{Convergence curves at RX-SNR = 10 dB for UEs moving at 5 km/h ($M=16$, $K=8$) under 4-bit RIS phase quantization. All algorithms exploit all RIS links. SCRPA is evaluated with $S=5$, $10$, and $15$, and results show the mean and standard deviation over 50 channel realizations.} 
\vspace{-9pt}
\label{fig:WSR_CONV_ALL}
\end{figure}

\subsection{Empirical Evaluation of WSR Optimization}
\label{subsec:wsr_eval}

All WSR results are evaluated on the true instantaneous, geometry-consistent mmWave channels, while the proposed method and all baselines rely only on imperfect and aged CSI for optimization.
Unless otherwise stated, the same channel realizations are shared across all compared methods to ensure a fair comparison.
The detailed dataset generation procedure uses a ray-tracing-based simulator as in Section~\ref{subsec:dataset}.

Results compare the proposed SCRPA framework with representative AO and manifold-based baselines widely used for RIS-assisted WSR maximization:

\vspace{2pt}
\noindent\textbf{AO--SPR}~\cite{LI2024WSRCRA}:
A deterministic AO algorithm for instantaneous WSR maximization with fixed CSI, where the BS precoder is updated via a WMMSE-based step and the RIS phases are optimized by element-wise sequential phase rotation (SPR).
For multi-RIS systems, the updates are applied by alternately optimizing the phase shifts of different RIS panels.

\noindent\textbf{RMO--LS}~\cite{NIU2022}:
A Riemannian manifold optimization baseline under deterministic CSI, in which the RIS phases are updated using Riemannian gradient descent with retraction and line search, and the BS precoder is obtained via a power-constrained WMMSE update.
The method extends to multi-RIS settings by alternating the manifold updates across RIS panels, using sample-average gradients when channel samples are available.

\noindent\textbf{AO--RMCG}~\cite{LI2020}:
A deterministic AO baseline for instantaneous WSR maximization assuming perfect CSI, where the BS precoder is updated via a Lagrangian-based WMMSE solution and the RIS phases are jointly refined using a Riemannian conjugate-gradient (RMCG) method on the unit-modulus manifold.

\noindent\textbf{WMMSE--GRAD}~\cite{WSR_MUMISO_GUO2020}:
A gradient-based baseline for WSR maximization under imperfect CSI, where the BS precoder is updated by (stochastic) gradient descent on a WMMSE-equivalent objective using available channel samples, while performance is evaluated on the true instantaneous channel.

\noindent \textbf{Strong LP Benchmark}: 
A deterministic AO benchmark under perfect CSIT, where the linear BS precoder is updated via a weighted WMMSE step and each RIS panel is optimized by directly maximizing the quadratic WMMSE surrogate with respect to the unit-modulus phase coefficients.
The RIS phase update is handled via an augmented semidefinite relaxation (SDR) with Gaussian randomization~\cite{LUO2010SDR}, yielding a strong upper-bound--style reference without constituting a global capacity upper bound.

\begin{figure*}[t!]
\captionsetup[subfloat]{farskip=2pt}
\vspace{-7pt}
\centering
        \subfloat[$M=8$, $K=4$, $v=5$~km/h]{\includegraphics[width=.33\linewidth]{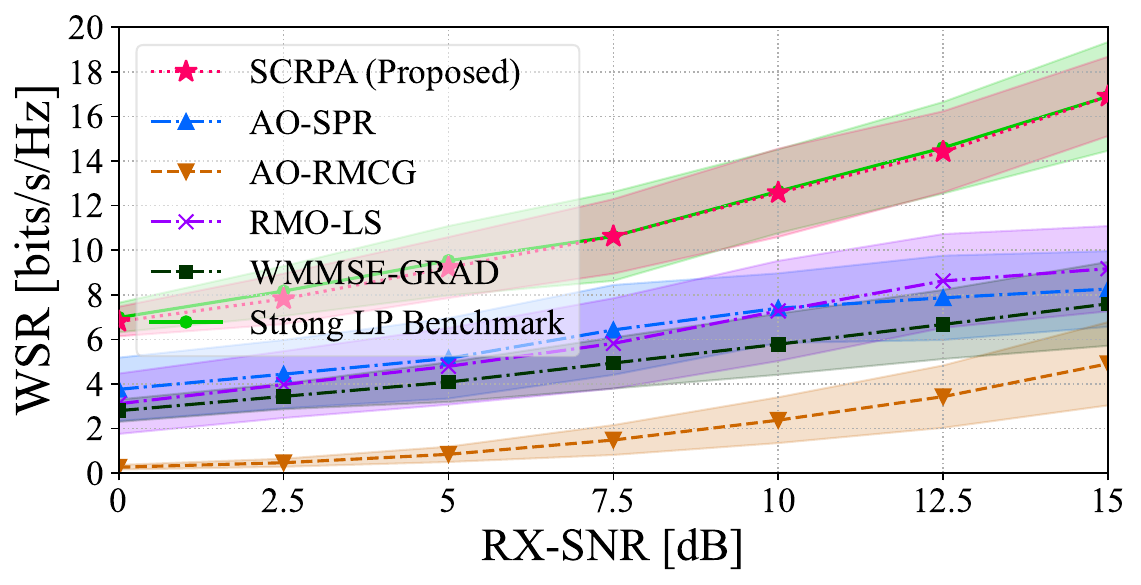}}
        \subfloat[$M=16$, $K=8$, $v=5$~km/h]{\includegraphics[width=.33\linewidth]{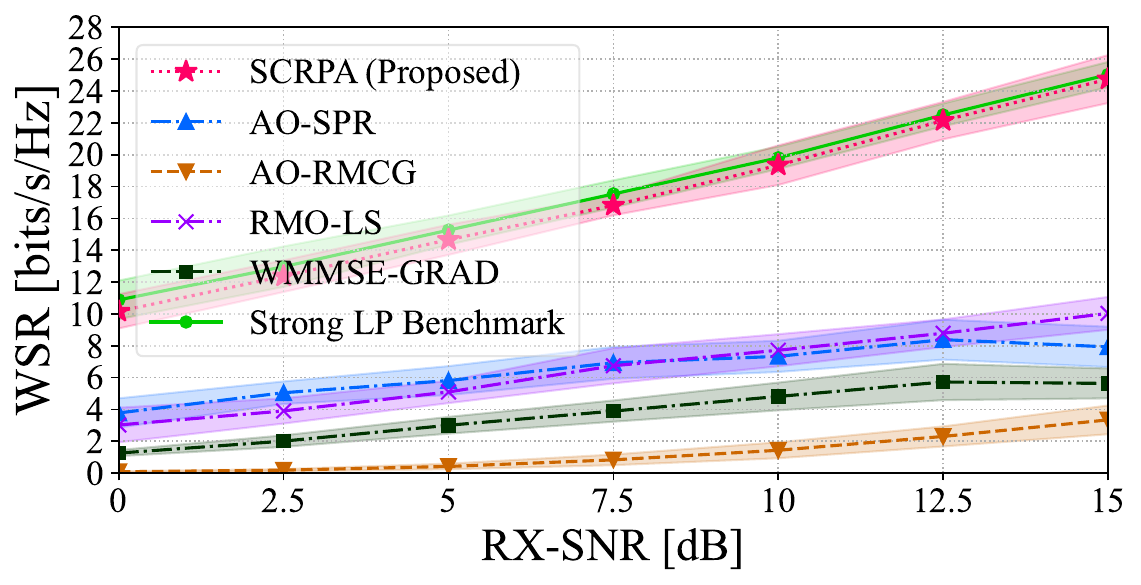}}
        \subfloat[$M=32$, $K=16$, $v=5$~km/h]{\includegraphics[width=.33\linewidth]{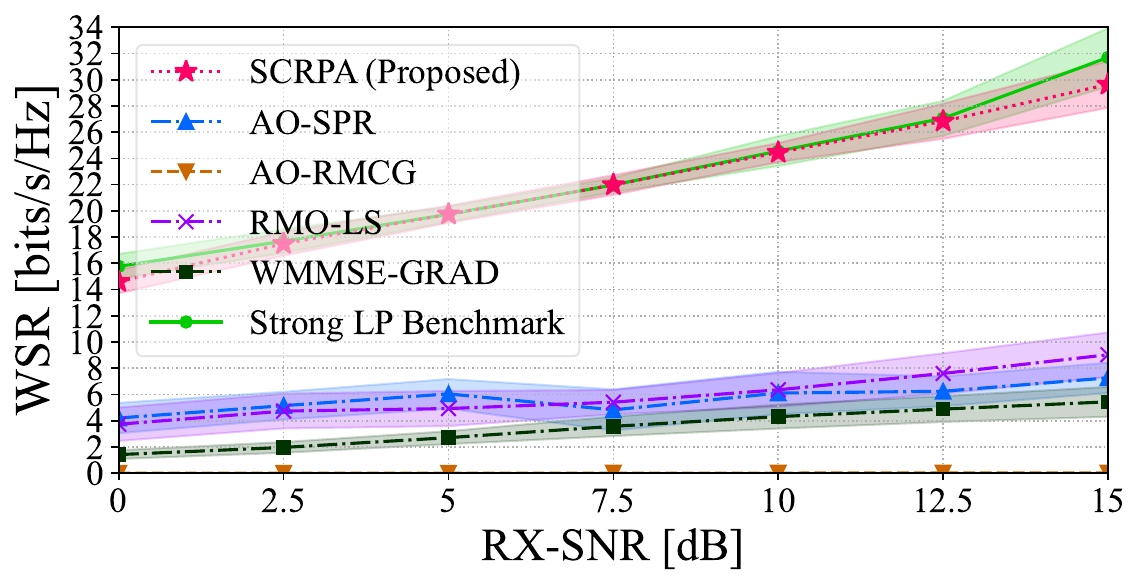}}\\
        \subfloat[$M=8$, $K=4$, $v=15$~km/h]{\includegraphics[width=.33\linewidth]{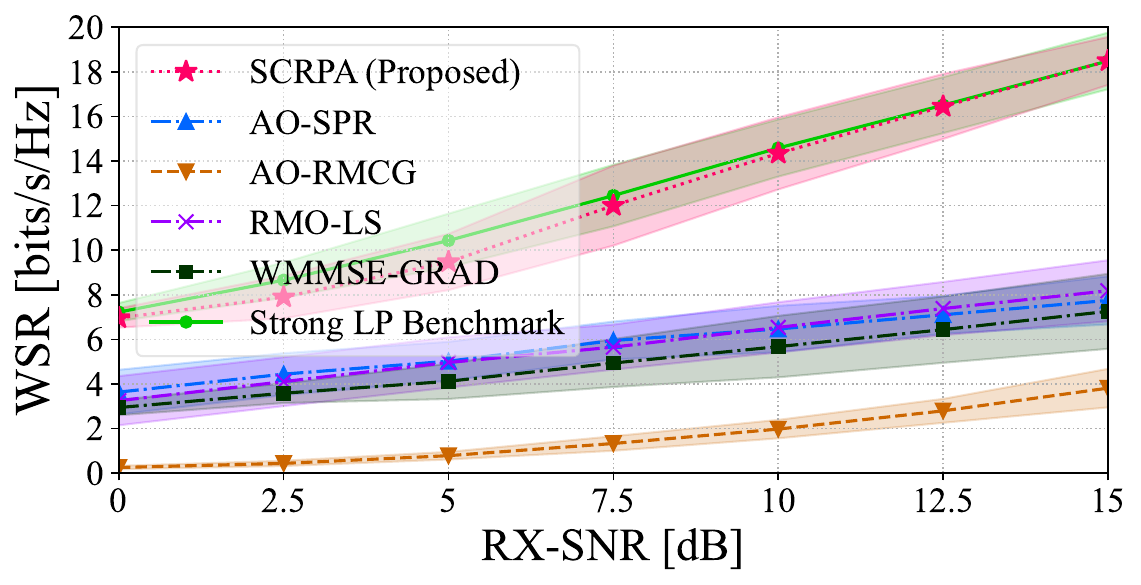}}
        \subfloat[$M=16$, $K=8$, $v=15$~km/h]{\includegraphics[width=.33\linewidth]{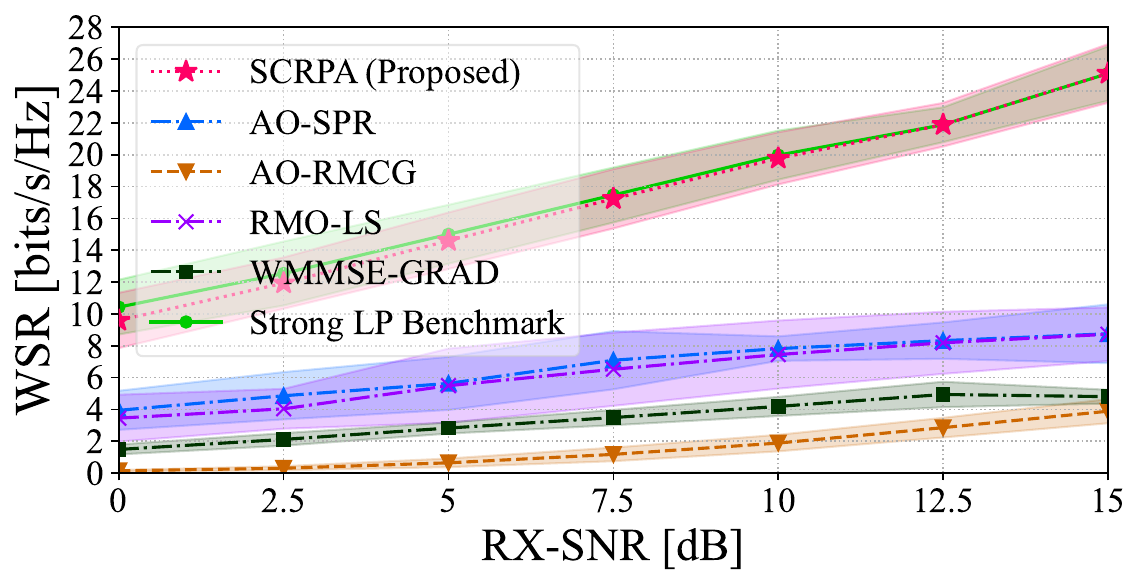}}
        \subfloat[$M=32$, $K=16$, $v=15$~km/h]{\includegraphics[width=.33\linewidth]{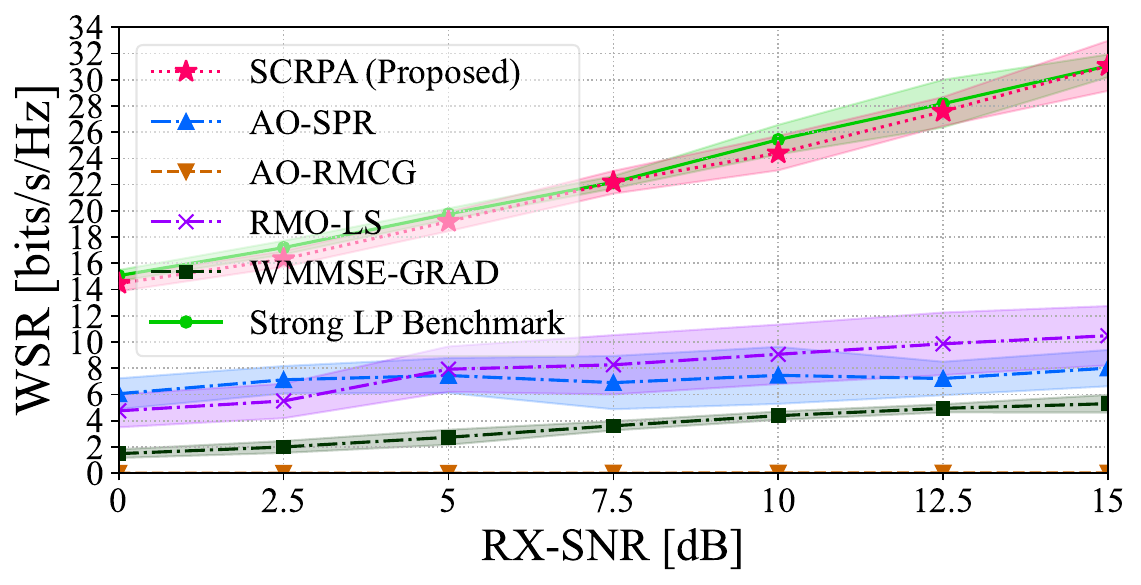}}
\vspace{-3pt}
\caption{WSR versus RX-SNR for UEs moving at 5 km/h and 15 km/h under different $(M,K)$ configurations, with 4-bit RIS quantization and 30\% per-link blockage. Results use $S=5$ samples. SCRPA reports end-to-end performance, including per-RIS blockage detection using an $m$-sequence of length $\ell=127$, while other algorithms assume all links are available. Solid lines show the mean over 50 channel realizations, with shaded regions indicating one standard deviation.} 
\vspace{-2pt}
\label{fig:WSR_SNR_E2E}
\end{figure*}

\begin{table*}[t!]
\centering
\caption{Average computational FLOPs per UE required for blockage detection, and the total FLOPs required for each algorithm to reach 99\% of its convergence value, under different configurations of the number of $M$, $K$, and $\ell$. All results are averaged over 50 independent channel realizations.}
\label{tab:flops}
\adjustbox{width=0.93\linewidth}{
\arrayrulecolor{lightgray}
\begin{tabular}{cc|cc|
 >{\centering\arraybackslash}p{1.6cm}
 >{\centering\arraybackslash}p{1.6cm}
 >{\centering\arraybackslash}p{1.6cm}
 >{\centering\arraybackslash}p{1.6cm}
 >{\centering\arraybackslash}p{1.9cm}}

\arrayrulecolor{black}
\toprule[1pt]
\shortstack{\textbf{\#RIS}\\\textbf{($M$)}} &
\shortstack{\textbf{\#UE}\\\textbf{($K$)}} &
\shortstack{\textbf{m-seq. length}\\\textbf{($l$)}} &
\textbf{\shortstack{Detection\\FLOPs}} &
\textbf{\shortstack{SCRPA\\FLOPs}} &
\textbf{\shortstack{AO-SPR\\FLOPs}} &
\textbf{\shortstack{AO-RMCG\\FLOPs}} &
\textbf{\shortstack{RMO-LS\\FLOPs}} &
\textbf{\shortstack{WMMSE-GRAD\\FLOPs}} \\
\hline

\multirow{3}{*}{8}  & \multirow{3}{*}{4}  & 127 & 0.59M & \multirow{3}{*}{0.243G} & \multirow{3}{*}{0.072G} & \multirow{3}{*}{1.427G} & \multirow{3}{*}{3.776G} & \multirow{3}{*}{0.009G} \\
   & & 255 & 0.61M & & & & & \\
   & & 511 & 0.64M & & & & & \\
\hline

\multirow{3}{*}{16} & \multirow{3}{*}{8}  & 127 & 1.14M & \multirow{3}{*}{0.969G} & \multirow{3}{*}{0.174G} & \multirow{3}{*}{5.123G} & \multirow{3}{*}{24.088G} & \multirow{3}{*}{0.027G} \\
    & & 255 & 1.17M & & & & & \\
    & & 511 & 1.24M & & & & & \\
\hline

\multirow{3}{*}{32} & \multirow{3}{*}{16} & 127 & 2.24M & \multirow{3}{*}{3.867G} & \multirow{3}{*}{0.514G} & \multirow{3}{*}{19.714G} & \multirow{3}{*}{54.997G} & \multirow{3}{*}{0.093G} \\
    & & 255 & 2.31M & & & & & \\
    & & 511 & 2.44M & & & & & \\
\bottomrule[1pt]
\vspace{-10pt}
\end{tabular}
\arrayrulecolor{black}
}
\end{table*}

\vspace{0.3em}
Fig. \ref{fig:WSR_SNR_ALL} shows the WSR performance under user mobility corresponding to a pedestrian speed of 5 km/h.
The proposed SCRPA effectively exploits temporally aged CSIT by constructing a channel prediction from a small number of recent channel observations (three samples) and explicitly modeling the residual uncertainty via the SAA framework.
By leveraging temporal correlation across consecutive channel realizations and incorporating the resulting estimation error covariance into the stochastic channel model in \eqref{eq:Hkstochastic_scrpa}, SCRPA maintains stable and high WSR performance in time-varying channels, whereas baseline algorithms based on deterministic instantaneous CSI suffer from severe performance degradation under mobility.
Although the beam coherence time at mmWave frequencies is typically longer than the CSI update interval, even small phase perturbations within a single update interval can significantly degrade coherent beamforming and RIS phase alignment due to the highly phase-sensitive nature of RIS-assisted channels.
By optimizing the BS precoder and RIS phases over a distribution of plausible channels within the CSI update interval rather than a single channel realization, SCRPA effectively mitigates this sensitivity, resulting in only marginal performance differences across different SAA sample sizes.
This indicates that a moderate Monte Carlo budget is sufficient to capture the dominant second-order statistics of the channel uncertainty and enable robust adaptation within practical CSI update intervals.

Fig. \ref{fig:WSR_CONV_ALL} illustrates the convergence behavior under the same system configuration.
SCRPA reaches 99\% of its converged WSR after approximately 386 iterations on average, whereas AO-SPR and RMO-LS, which achieve the second and third-best WSR performance, converge in 47 and 413 iterations, respectively.
While SCRPA requires more iterations than AO-SPR, this is expected since SCRPA optimizes a stochastic sample-average objective that explicitly accounts for CSI aging, in contrast to deterministic instantaneous-CSI-based baselines.
Importantly, SCRPA converges to a substantially higher WSR, demonstrating that the increased iteration count reflects the higher complexity and robustness of the underlying optimization problem rather than inefficient convergence.

\section{Numerical Experiments}
\label{sec:experiments}

\begin{figure}[h]
\centering
    \includegraphics[width=0.76\linewidth]{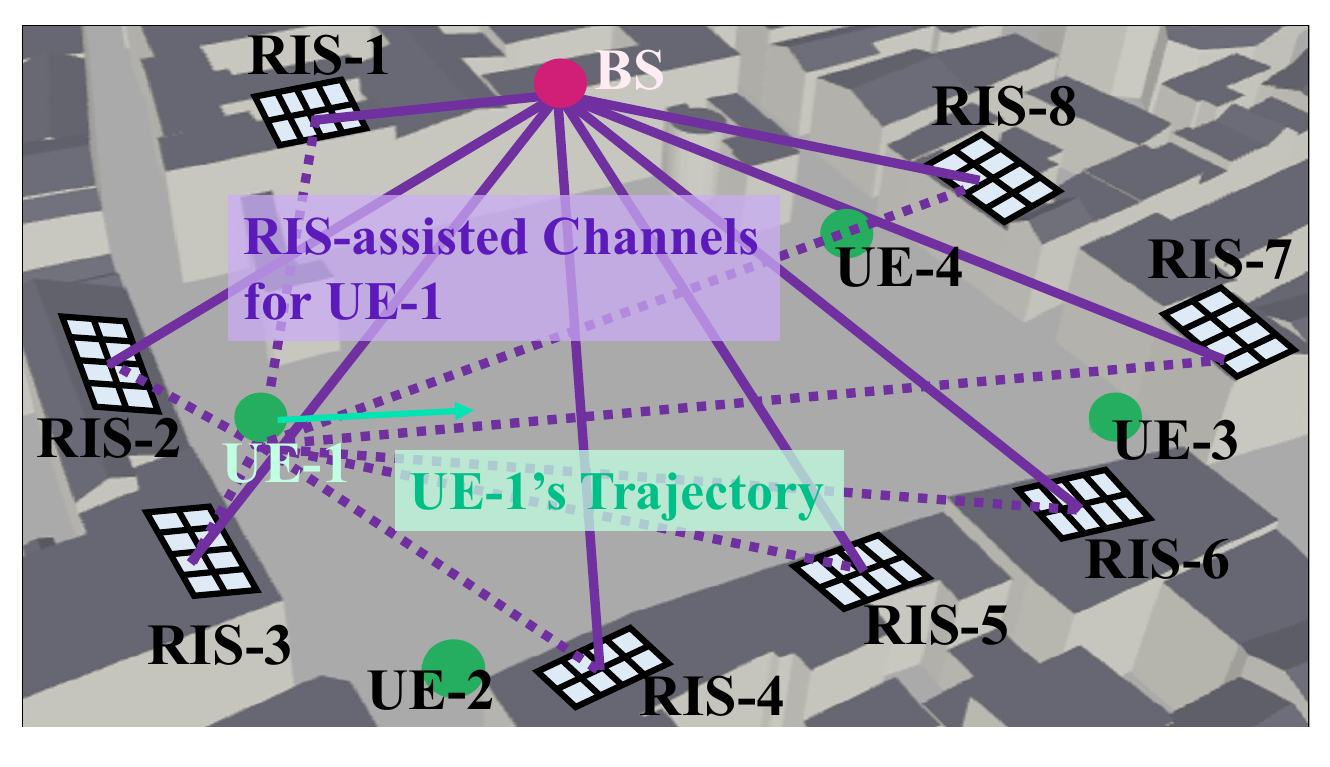}
    \caption{Ray-tracing–based channel generation in the Munich urban scene of NVIDIA Sionna, illustrating the placement of the BS, multiple RIS panels, and UEs (example with $M=8$ and $K=4$).}
    \label{fig:dataset}
    \vspace{-10pt}
\end{figure}

\begin{figure*}[t!]
\captionsetup[subfloat]{farskip=2pt}
\vspace{-7pt}
\centering
        \subfloat[$M=8$, $K=4$]{\includegraphics[width=.33\linewidth]{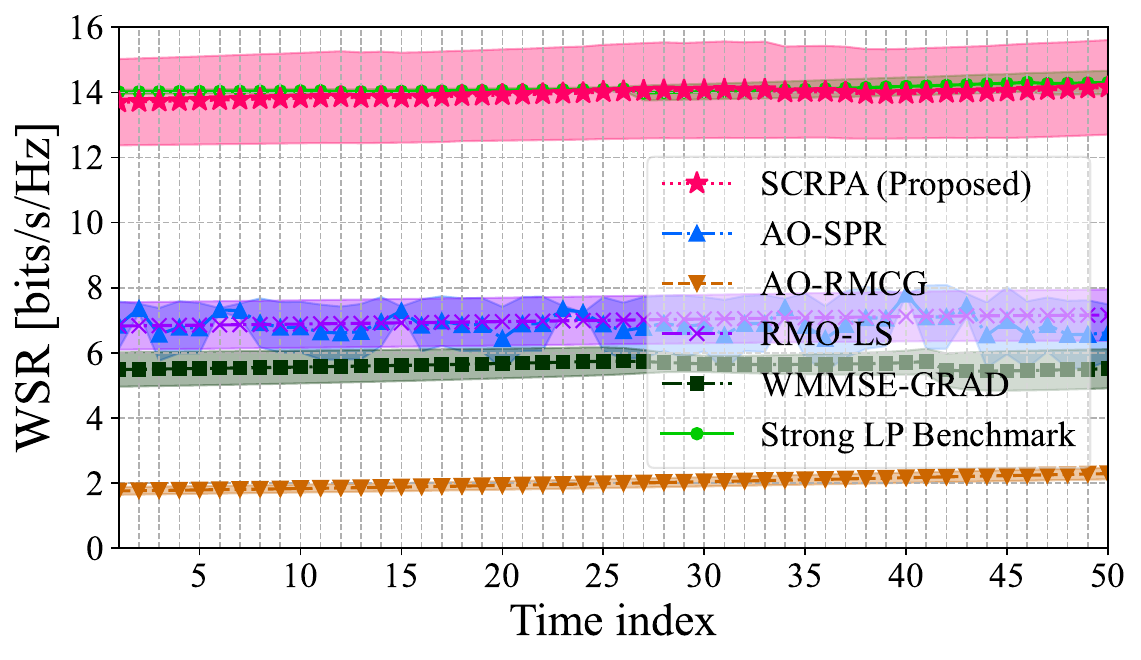}}
        \subfloat[$M=16$, $K=8$]{\includegraphics[width=.33\linewidth]{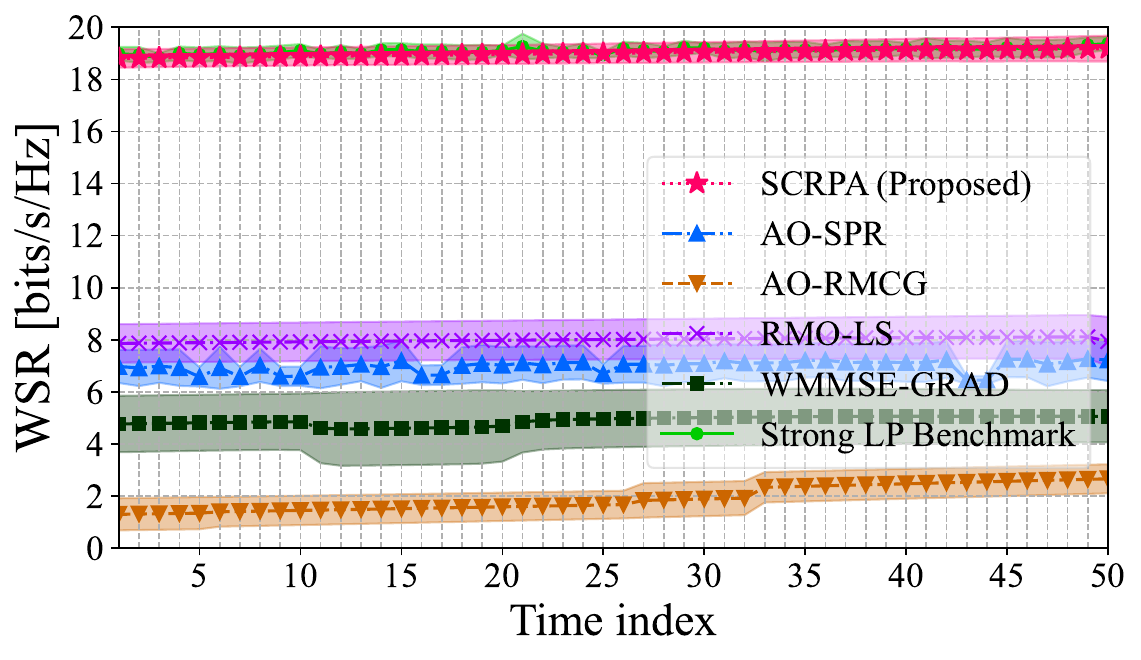}}
        \subfloat[$M=32$, $K=16$]{\includegraphics[width=.33\linewidth]{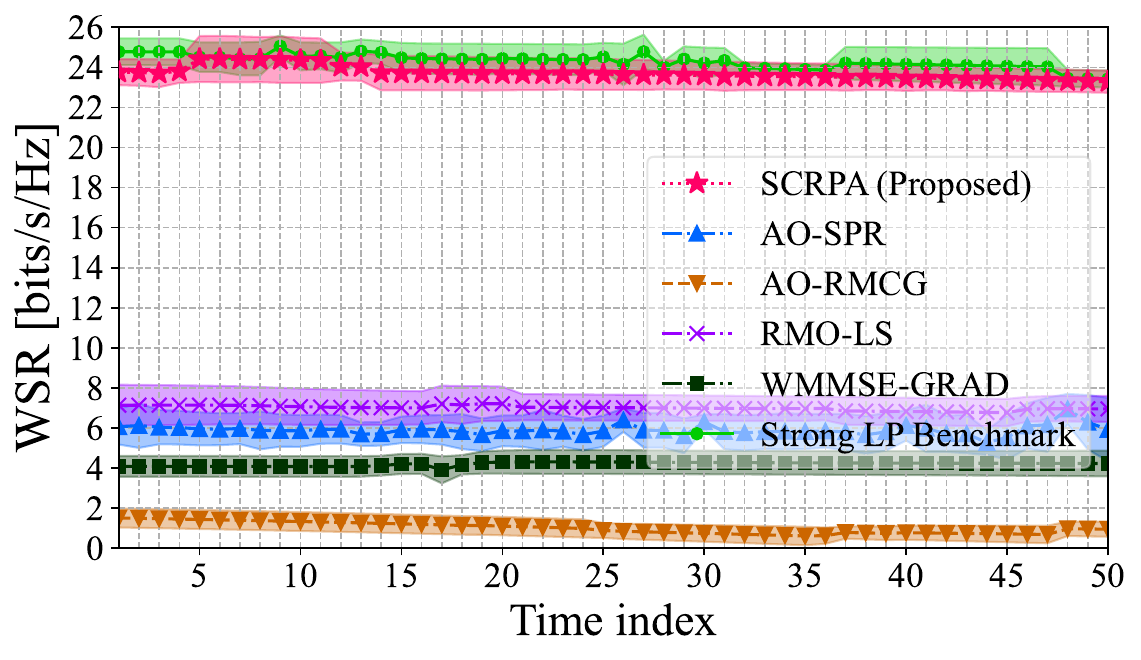}}
\vspace{-3pt}
\caption{For UEs moving at $v=10$ km/h, the WSR is evaluated at a time resolution of 5 ms. All algorithms optimize their transmission parameters sequentially using only past channel state information, and results show the mean and standard deviation over 50 channel realizations.} 
\vspace{-2pt}
\label{fig:WSR_MOB}
\end{figure*}

\begin{figure}[t]
\captionsetup[subfloat]{farskip=2pt}
\vspace{-7pt}
\centering
    \includegraphics[width=0.95\linewidth]{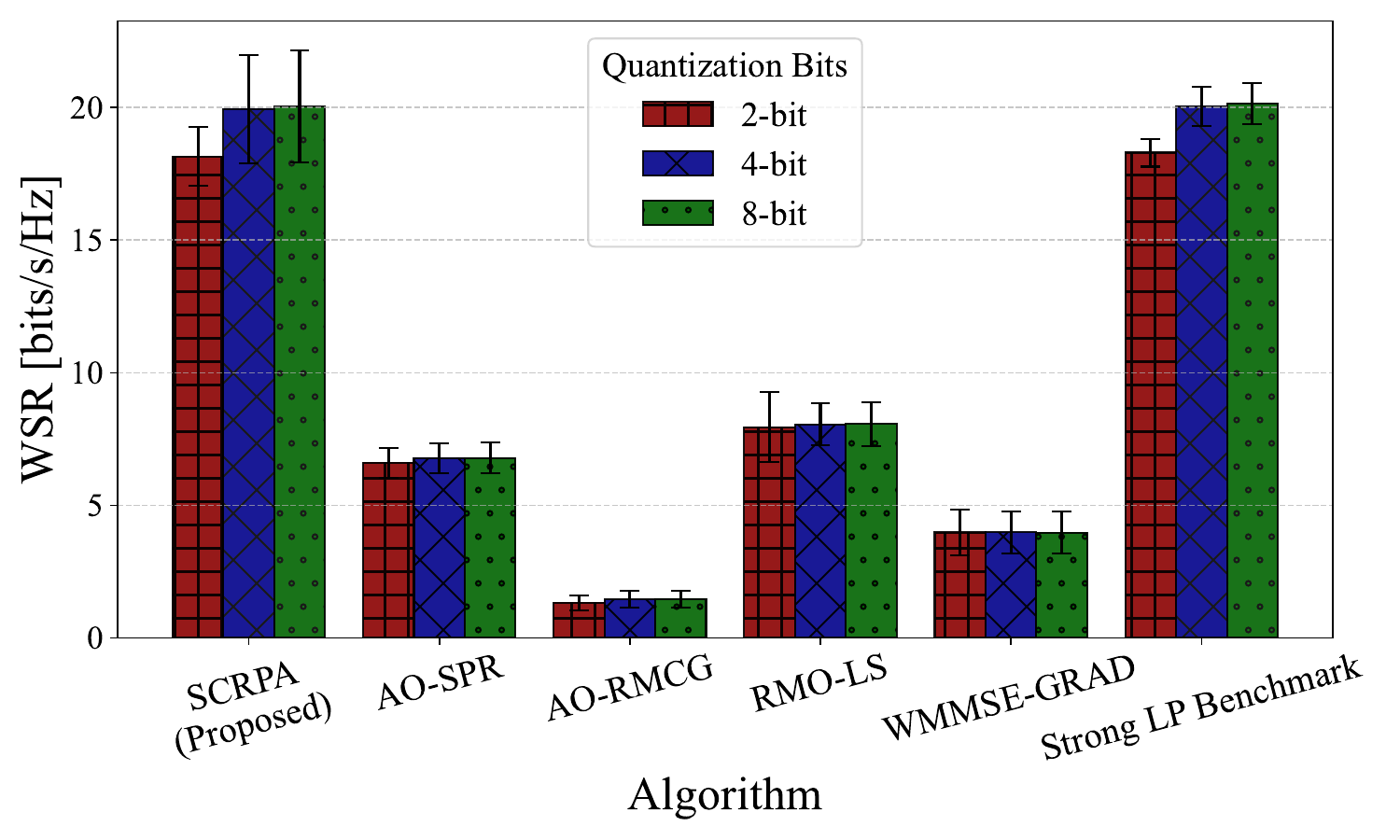}
\vspace{-3pt}
\caption{SCRPA performance at RX-SNR = 10 dB for 10 km/h mobility ($M=16$, $K=8$) under 2, 4, and 8-bit RIS quantization. Results show the mean WSR over 50 independent channel realizations, with error bars indicating one standard deviation.} 
\vspace{-2pt}
\label{fig:QUANT}
\end{figure}

\subsection{Dataset}
\label{subsec:dataset}

All datasets are generated using the ray-tracing-based Munich urban scene provided by the Sionna simulator~\cite{NVIDIA}, enabling geometry-consistent mmWave propagation with mobility, blockage, and RIS-assisted reflections.
The overall scene layout and network geometry are illustrated in Fig.~\ref{fig:dataset}.

The considered deployment follows a typical urban microcell (UMi) scenario, as commonly adopted in 3GPP mmWave evaluations~\cite{TS38827}, where the cell radius spans from several tens to a few hundred meters.
The BS is placed at a fixed location with coordinates $(8.5, 21, 27)$ and is equipped with a planar antenna array with $N_t$ elements.
Unless stated otherwise, the BS array is configured as a uniform planar array (UPA), with dimensions chosen according to the total number of antennas ($4\times4$ for $N_t=16$, $4\times8$ for $N_t=32$, and $8\times8$ for $N_t=64$).
A total of $M$ RIS panels are deterministically distributed along the perimeter of a rectangular region defined by the vertices $(-30,80,26)$, $(120,40,26)$, $(110,110,26)$, and $(-30,140,26)$.
In addition, one RIS panel is deployed on the side opposite to the BS location to ensure coverage in the facing direction.
For $M=8$, one RIS is placed at each vertex and one at the midpoint of each edge.
for $M=16$, one RIS is placed at each vertex and three RISs are sampled equidistantly along each edge;
and for $M=32$, one RIS is placed at each vertex and seven RISs are sampled equidistantly along each edge.
Each RIS consists of $N_i$ passive reflecting elements arranged as a planar array, with array dimensions matched to the total element count (e.g., $4\times4$, $4\times8$, or $8\times8$), and supports per-element phase control with finite resolution.

UEs are initialized uniformly at random within a central urban microcell area bounded by $(-20, 70, 1.5)$, $(90, 40, 1.5)$, $(-10, 130, 1.5)$, and $(100, 100, 1.5)$, corresponding to an area on the order of $10^2$ meters.
This spatial scale is consistent with standard UMi assumptions at mmWave frequencies, where BS–UE distances typically range from tens to a couple of hundred meters.
Each UE is equipped with a planar receive array of $N_{r,k}$ elements (typically $N_{r,k}\ll N_t$) and moves along a straight-line trajectory with a randomly selected direction, capturing mobility-induced channel variations within the UMi environment.
Channel realizations are generated every $\Delta t=5$~ms, corresponding to a periodic CSI update interval configured via the CSI resource/report periodicity-and-offset options defined in 3GPP TS 38.331~\cite{TS38331} and TS 38.214~\cite{TS38214}.
This choice aligns with CSI acquisition and beam-management time scales in 3GPP NR FR2 systems and ensures that multiple channel samples fall within a single beam-coherence interval.
UE speeds satisfy $v\in\{5,10,15\}$~km/h, representative of pedestrian and low-mobility mmWave scenarios.
At these speeds, dominant beam directions remain stable across several samples, while small-scale fading and phase variations are non-negligible.

\subsection{End-to-End Operation under Mobility}

Fig.~\ref{fig:WSR_SNR_E2E} presents the end-to-end performance when the proposed blockage detection in Section~\ref{sec:blockage_detection} and the SCRPA-based WSR optimization in Section~\ref{sec:scrpa} are jointly applied under a 30\% per-link blockage probability.
Across all considered pedestrian speeds and $(M,K)$ configurations, SCRPA consistently achieves substantially higher WSR than the baseline methods.
Moreover, its performance closely approaches that of the Strong LP Benchmark, demonstrating the effectiveness of the proposed end-to-end design.

Fig.~\ref{fig:WSR_MOB} further evaluates the proposed framework under continuous user mobility.
Specifically, users move according to the prescribed trajectories, and the end-to-end algorithm is executed sequentially at a time resolution of 5 ms.
The figure reports the average WSR at each time instant.
The results indicate that, as long as the detection--optimization loop can be completed within 5 ms, the proposed method maintains a stable and high WSR despite time-varying channels, confirming its suitability for mobility-aware
mmWave operation.

Table~\ref{tab:flops} summarizes the computational complexity in terms of the floating-point operations (FLOPs) required for blockage detection and for the WSR optimization until convergence.
Among all compared algorithms, SCRPA requires the third smallest number of FLOPs to converge, while achieving significantly higher WSR than the lower-complexity baseline (WMMSE-GRAD, AO-SPR).
When both performance and complexity are considered jointly, SCRPA therefore offers the most favorable end-to-end trade-off.

For the largest evaluated configuration with $M=32$ RIS panels and $K=16$ UEs, the total computational cost of SCRPA is approximately $3.869$~GFLOPs.
To complete the proposed end-to-end processing within a $5$~ms budget, a sustained FP32 throughput of approximately $0.774$~TFLOPs/s is required.
This throughput requirement lies well within the capabilities of contemporary general-purpose computing platforms, including modern multi-core CPUs, GPUs, and dedicated accelerators.
In practice, sustained FP32 throughputs on the order of $1$~TFLOPs/s are achievable via parallel execution and optimized linear algebra kernels, even when accounting for implementation overheads.
As a result, end-to-end execution times on the order of a few milliseconds are practically attainable under realistic processing conditions.
In comparison, the physical signal propagation delay over typical urban microcell (UMi) distances of $100$--$200$~m is approximately $0.3$--$0.7~\mu$s, given the speed of light.
As a result, the sensing-phase airtime and propagation delay in \eqref{eq:Tsense_def} are negligible compared to the computational latency.
Therefore, the overall detect--optimize--transmit pipeline is dominated by computation rather than propagation delay and can be reliably executed within the CSI update interval, thereby satisfying the timing constraint in \eqref{eq:timing_const} and validating the practical feasibility of the proposed end-to-end framework.

\subsection{Robustness and Scalability}

Fig.~\ref{fig:WSR_SNR_E2E} and Fig.~\ref{fig:WSR_MOB} demonstrate the scalability of the proposed SCRPA framework across a wide range of RIS and UE configurations.
As the numbers of RIS panels and UEs increase, the dimensionality of the effective channels and the associated optimization variables grows rapidly, leading to a substantial increase in computational complexity.
Nevertheless, the proposed algorithm maintains stable convergence behavior and consistent WSR gains, indicating that SCRPA can be applied in a scalable manner without degradation in performance.

Furthermore, Fig.~\ref{fig:QUANT} evaluates the robustness of the proposed scheme under practical RIS phase quantization constraints.
The results show that SCRPA exhibits limited performance sensitivity to phase quantization, with comparable WSR performance for 2, 4, and 8-bit resolutions.
In particular, the performance gap between 4-bit and 8-bit quantization remains within approximately 0.3\%, demonstrating strong robustness to coarse quantization.
Across all quantization levels, the proposed method consistently outperforms the baseline schemes by leveraging explicit blockage detection and predictive channel modeling within a stochastic optimization framework, and achieves performance close to the Strong LP Benchmark.

These results confirm that SCRPA is resilient to both system scaling and hardware-imposed quantization constraints, highlighting its practical viability for real-world multi-RIS deployments.

\section{Summary}
\label{sec:conclusion}

This paper presented an end-to-end framework for weighted sum-rate (WSR) maximization in mmWave multi-RIS downlink systems that explicitly addresses user mobility and dynamic blockage under time-varying and imperfect CSI.
By moving beyond idealized assumptions of static or perfectly known channels, the proposed approach bridges the gap between theoretical multi-RIS performance gains and practical implementation constraints.

A standard-compliant per-RIS blockage detection protocol based on indexed $m$-sequences was developed to reliably identify active RIS panels with negligible overhead.
Leveraging the detected active set, we formulated a robust WSR maximization problem that accounts for CSI aging and feedback delay, and proposed SCRPA, a stochastic block majorization--minimization framework enabling closed-form BS precoder updates and hardware-constrained RIS phase optimization.

Analytical complexity and timing evaluations verified that the proposed detection--optimization--transmission loop satisfies practical CSI update interval requirements, while ray-tracing-based simulations demonstrated significant spectral efficiency gains over representative baselines under CSI uncertainty and RIS phase quantization.
The results further confirm favorable scalability with respect to the number of RIS panels and users, supporting dense multi-RIS deployments.

Future work includes wideband OFDM extensions with frequency-selective channels and blockage effects, tighter integration with 3GPP NR beam management, experimental validation on hardware testbeds, and the integration of the canonical multiuser transmission framework with RIS phase optimization.

\appendices

\section{Proof of Theorem~\ref{thm:conv_scrpa}}
\label{appendix:proof_conv_scrpa}

We prove the convergence properties of SCRPA for the fixed-ensemble
SAA WMMSE problem in \eqref{eq:SAA_obj_scrpa}.
The proof follows standard block MM and block coordinate descent arguments \cite{RAZAVIYAYN2013BMM}.

\subsection{Preliminaries}

Fix an outer iteration index $t$ and the Monte Carlo ensemble $\{\mathbf{H}_{k,\mathrm{eff}}^{(s,t)}\}_{k,s}$ generated by \eqref{eq:Hkstochastic_scrpa}.
For notational simplicity, the dependence on $(s,t)$ is omitted.
Collect all optimization variables as
\[
\mathcal{V}
\triangleq
\Big(
\mathbf{F}, \{u_i\},
\{\mathbf{U}_k^{(s)}, \mathbf{W}_k^{(s)}\}_{k,s}
\Big),
\]
and define the fixed-ensemble SAA WMMSE objective
\begin{equation}
\label{eq:J_def_app}
J(\mathcal{V})
=
\frac{1}{S}
\sum_{s=1}^S
\sum_{k=1}^K
\Big(
\mathrm{Tr}(\mathbf{W}_k^{(s)}\mathbf{E}_k^{(s)})
-
\log\det \mathbf{W}_k^{(s)}
\Big).
\end{equation}

The function $J(\mathcal{V})$ is continuous and bounded below over the feasible set.
The BS precoder feasible set $\{\mathbf{F} : \|\mathbf{F}\|_F^2 \le P\}$ is compact, and the RIS feasible set is compact for continuous phases or finite for quantized phases.

\subsection{Monotonic Descent of Each Block Update}

We consider one full SCRPA iteration indexed by $r$, consisting of successive updates of the receiver/weight blocks, the BS precoder, and the RIS phases.

\paragraph*{Receiver and weight updates}
For fixed $(\mathbf{F}, \{u_i\})$, the objective $J(\mathcal{V})$ is separable across $(k,s)$ in the blocks $\{\mathbf{U}_k^{(s)}, \mathbf{W}_k^{(s)}\}$.
The MMSE receiver update in \eqref{eq:update_u} and the optimal weight update $\mathbf{W}_k^{(s)} = (\mathbf{E}_k^{(s)})^{-1}$ in \eqref{eq:update_w} jointly minimize each block subproblem, yielding
\[
J(\mathcal{V}^{r,1}) \le J(\mathcal{V}^{r}),
\]
where $\mathcal{V}^{r,1}$ denotes the variable collection after this block update.

\paragraph*{BS precoder update}
For fixed $\{u_i\}$ and $\{\mathbf{U}_k^{(s)}, \mathbf{W}_k^{(s)}\}$, $J(\mathcal{V})$ is a convex quadratic function of $\mathbf{F}$ under the power constraint.
The update in \eqref{eq:update_F_nofair} yields the global minimizer, implying
\[
J(\mathcal{V}^{r,2}) \le J(\mathcal{V}^{r,1}),
\]
where $\mathcal{V}^{r,2}$ denotes the variables after the precoder update.

\paragraph*{RIS phase updates (continuous phases)}
Fix all variables except $u_i$.
SCRPA constructs a quadratic majorizer
\begin{equation}
\label{eq:MM_majorizer_app}
\begin{aligned}
J(u_i;\cdot)
\;\le\;
M_i(u_i; u_i^{r})
&=
J(u_i^{r};\cdot)
+
\mathrm{Re}\!\left\{ (g_i^{r})^H (u_i - u_i^{r}) \right\} \\
&\quad
+
\frac{L_i}{2}\,
\|u_i - u_i^{r}\|_2^2 ,
\end{aligned}
\end{equation}
where $g_i^{r}$ is the gradient of $J$ with respect to $u_i$ and $L_i$ is chosen (e.g., via backtracking) to ensure majorization.
Minimizing $M_i(u_i; u_i^{r})$ under the unit-modulus constraint $|[u_i]_m|=1$ yields the closed-form update in \eqref{eq:update_closed}.
By the MM principle,
\[
J(u_i^{r+1};\cdot) \le J(u_i^{r};\cdot).
\]
Sequentially updating all RIS panels yields
\[
J(\mathcal{V}^{r+1}) \le J(\mathcal{V}^{r,2}).
\]

\paragraph*{RIS phase updates (finite-resolution phases)}
When $[u_i]_m \in \mathcal{U}_\beta$, the continuous-phase update is projected elementwise onto the discrete feasible set.
This projection yields the optimal solution of the discrete MM subproblem and ensures a monotonic decrease of the MM surrogate.
Since the feasible set is finite, the resulting sequence converges to a discrete block-stationary point.

\subsection{Convergence}

\paragraph*{Continuous phases}
Combining the above block updates, one full SCRPA iteration satisfies
\[
J(\mathcal{V}^{r+1}) \le J(\mathcal{V}^{r}), \quad \forall r.
\]
Since $J$ is bounded below, the sequence $\{J(\mathcal{V}^{r})\}$ converges.
By standard block-MM convergence results \cite{RAZAVIYAYN2013BMM}, every accumulation point of $\{\mathcal{V}^{r}\}$ is block-stationary for the fixed-ensemble SAA problem.

\paragraph*{Finite-resolution phases}
Under quantized RIS phases, SCRPA converges to a discrete block-stationary
solution in the sense that no single block update can further decrease the MM
surrogate.

This completes the proof of Theorem~\ref{thm:conv_scrpa}.
\hfill$\square$

\bibliographystyle{IEEEtran}
\bibliography{ref}

\end{document}